\documentclass[11pt]{article}
\pdfoutput=1 
\usepackage{jheppub}
\usepackage[utf8]{inputenc}
\usepackage{amsmath,amsthm}
\usepackage{cleveref}
\usepackage{graphicx,ulem}
\usepackage{caption}
\usepackage{subcaption}
\usepackage{hyperref}
\usepackage[T1]{fontenc}

\usepackage[table]{xcolor}
\usepackage{amssymb}
\usepackage{pifont}

\newcommand{\be}{\begin{equation}}
\newcommand{\ee}{\end{equation}}
\newcommand{\bea}{\begin{eqnarray}}
\newcommand{\eea}{\end{eqnarray}}

\newtheorem{Ass}{Assumption}

\crefname{Ass}{Assumption}{Assumptions}

\makeatletter
\gdef\@fpheader{}
\makeatother

\begin{document}
\title{\boldmath The Algebraic Structure Underlying Pole-Skipping Points}

\author[a]{Zhenkang Lu,}
\author[a]{Cheng Ran,}
\author[a,b]{Shao-Feng Wu,}

\affiliation[a]{Department of Physics, Shanghai University, Shanghai, 200444, China}
\affiliation[b]{Center for Gravitation and Cosmology, Yangzhou University, Yangzhou, 225009, China}

\emailAdd{zhenkanglu9@gmail.com}
\emailAdd{r\_cheng@shu.edu.cn}
\emailAdd{sfwu@shu.edu.cn}

\abstract{The holographic Green's function becomes ambiguous, taking the indeterminate form `$0/0$', at an infinite set of special frequencies and momenta known as ``pole-skipping points''. In this work, we propose that these pole-skipping points can be used to reconstruct both the interior and exterior geometry of a static, planar-symmetric black hole in the bulk. The entire reconstruction procedure is fully analytical and only involves solving a system of linear equations. We demonstrate its effectiveness across various backgrounds, including the BTZ black hole, its $T\bar{T}$-deformed counterparts, as well as geometries with Lifshitz scaling and hyperscaling-violation. Within this framework, other geometric quantities, such as the vacuum Einstein equations, can also be reinterpreted directly in terms of pole-skipping data. Moreover, our approach reveals a hidden algebraic structure governing the pole-skipping points of Klein-Gordon equations of the form $(\nabla^{2} + V(r))\phi(r) = 0$: only a subset of these points is independent, while the remainder is constrained by an equal number of homogeneous polynomial identities in the pole-skipping momenta. These identities are universal, as confirmed by their validity across a broad class of bulk geometries with varying dimensionality, boundary asymptotics, and perturbation modes.}

\maketitle


\section{Introduction}

The Anti-de Sitter/Conformal Field Theory (AdS/CFT) correspondence, or holography \cite{maldacena_1999_first_Holography, Gubser_1998_holography, witten_1998_Holography_2}, asserts that quantum gravity in an asymptotically AdS spacetime is dual to a strongly coupled conformal field theory (CFT) defined on its boundary. A central challenge in this framework is bulk reconstruction: the task of recovering the bulk spacetime geometry and its dynamical fields purely from data in the boundary quantum field theory (QFT). 

Broadly speaking, bulk reconstruction falls into two categories: bulk operator reconstruction and bulk metric reconstruction. The former seeks to identify local bulk operators using boundary information, while the latter aims to deduce the gravitational background geometry from the boundary QFT.

In the domain of bulk operator reconstruction, several influential approaches have been developed.\footnote{For a comprehensive review, see \cite{Harlow_2018_bulk_op_reconstruction_review}.} One of the most prominent is the Hamilton-Kabat-Lifschytz-Lowe method \cite{hamilton_2007_HKLL_Reconstruction_1, hamilton_2006_HKLL_Reconstruction_2, hamilton_2006_HKLL_Reconstruction_3}, which reconstructs local bulk operators in AdS directly from their dual CFT boundary operators. Another major framework is entanglement wedge reconstruction \cite{czech_2012_Entanglement_wedge_reconstruction_6, headrick_2014_Entanglement_wedge_reconstruction_1, wall_2014_entanglement_wedge_reconstruction_3, dong_2016_Entanglement_wedge_reconstruction_4, jafferis_2016_entanglement_wedge_reconstruction_5, cotler_2019_Entanglement_wedge_reconstruction_2, penington_2020_Bulk_reconstruction_Entanglement_Wedge_7}, which is applicable to more general bulk geometries. It enables the reconstruction of local operators within the entanglement wedge: the region bounded by a chosen boundary subregion and its associated Ryu-Takayanagi surface \cite{ryu_2006_rt_formula_original}. Additional techniques for bulk operator reconstruction can be found in \cite{faulkner_2019_Bulk_reconstruction_modular, grimm_2022_Bulk_reconstruction_moduli_space, nebabu_2024_Bulk_reconstruction_other_generalized_free_fields}.

This paper focuses instead on bulk metric reconstruction, which has also seen the development of numerous innovative approaches. Many are rooted in the idea that spacetime emerges from entanglement \cite{maldacena_2003_extend_AdS_and_Two_entangled_CFT, ryu_2006_rt_formula_original, raamsdonk_2010_build_up_Spacetime_entanglement, swingle_2012_entanglement_renormalization_geometry, maldacena_2013_ER_equal_to_EPR}, and thus aim to reconstruct the bulk geometry from boundary entanglement measurements \cite{bilson_2011_Entanglement_reconstruct_bulk_metric_2, czech_2014_Entanglement_reconstruct_bulk_metric_3, roy_2018_entanglement_modularHamiltonian_bulk_metric_reconstruction, bao_2019_Entanglement_reconstruct_bulk_metric_1, jokela_2023_lattice_entanglement_entropy_reconstruct_bulk_metric}. Other boundary quantities used for reconstruction include complexity \cite{hashimoto_2021_complexity_bulk_metric_1, xu_2023_Entanglement_reconstruct_bulk_metric_ML_1}, Wilson loops \cite{hashimoto_2021_Wilson_loops_bulk_reconstruction_metric_1}, light-cone cuts \cite{engelhardt_2017_other_Light_cone_Metric_reconstruction, engelhardt_2017_Other_light_cone_Metric_reconstruction_2, hernandez-cuenca_2020_Other_light_cone_Metric_reconstruction_3}, celestial multipoles \cite{compere_2022_Bulk_metric_reconstruction_celestial_multipoles}, and various features of boundary correlators \cite{deharo_2001_other_T_munv_Metric_reconstruction, hammersley_2006_Bulk_reconstruction_null_geodesic, bilson_2008_Correlation_function_bulk_metric_1, yang_2023_Greenfunction_to_metric_analytical}. In addition, an increasing number of studies have employed machine learning techniques to explore the emergence of bulk spacetime \cite{You_2017_ML_Random_network, Hashimoto_2018_AdS/DL_original, Dong_2018_ML_network,  Hashimoto_2018_ML_holographic_QCD, Hu_2019_ML_renormalization, Hashimoto_2019_ML_Boltzmann,  Han_2019_ML_bulk_matrice_model, Tan_2019_DL_charge_black_hole, Akutagawa_2020_DL_AdS_QCD, yan_2020_Metric_reconstruction_shear_viscosity, Hashimoto_2020_DL_AdS_QCD, Hashimoto_2021_ML_Dilaton_potential_QCD, Lam_2021_ML_multi_EE, Park_2022_DL_EE, Hashimoto_2022_ML_Dilaton_potential_QCD_2, Li_2022_ML_BH_conductivity, Ahn_2024_DL_optical_conductivity, Chen_2024_ML_lattice_QCD, Gu_2024_ML_holo_dissipation, Bea_2024_ML_inverse_equ_of_state, Chen_2024_ML_QCD_data_metric, Ahn_2024_ML_entanglement_entropy, koji_2025_ML_Green_function_reconstruct, Ahn_2025_ML_T-linear}. 
 
Almost all of the aforementioned approaches involve tasks such as performing calculus, solving differential equations, or training neural networks. In contrast, we propose an analytical reconstruction method that relies solely on elementary arithmetic, involving nothing more than solving systems of linear equations. The only boundary input needed for this approach is a set of special locations, known as ``pole-skipping points'', where the Green's function becomes ambiguous.

Pole-skipping is a compelling phenomenon observed in holographic Green's functions in momentum space. It occurs at specific frequencies and momenta where the Green's function becomes ambiguous: taking the indeterminate form `$0/0$’. Typically, the pole-skipping frequencies lie at complex Matsubara frequencies,\footnote{For a derivation based on covariant expansion formalism, see \cite{wang_2022_Pole-skipping_gauge_bosonicfields, ning_2023_pole-skipping_gauge_fermionicfields}; an alternative approach related to algebraically special frequencies is provided in \cite{grozdanov_2024_Constraints_on_spectrum_infinte_product, Grozdanov_2025_4d_spectral_constraint_2}.} while the associated momenta depend on the specific details of different theories. This phenomenon originated in the study of quantum chaos, where the pole-skipping point of the energy-density Green’s function was shown to encode key chaotic data---namely, the Lyapunov exponent and butterfly velocity \cite{blake_2018_upper_Chaos_pole_skipping_1, grozdanov_2018_upper_Chaos_pole_skipping_0, blake_2018_Upper_Chaos_pole_skipping_2}. These chaos-related pole-skipping points universally appear in the upper-half complex frequency plane, particularly at $\omega_{c} = i2\pi T$. Subsequent developments have linked them to an emergent horizon symmetry \cite{knysh_2024_Horizon_symmetry_pole}, and reinterpreted them via a gravitational replica manifold that encodes late-time entanglement wedge \cite{Chua_2025_replica_pole_skipping}. Beyond quantum chaos, a countably infinite set of additional pole-skipping points, presumed to be unrelated to chaos, have been found in various Green’s functions \cite{blake_2020_lower_infinite_pole_skipping_1}. These points typically lie in the lower-half complex frequency plane.

From the perspective of bulk gravity, pole-skipping points originate from ambiguities in specifying ingoing boundary conditions at the black hole horizon, where two linearly independent ingoing solutions simultaneously satisfy the bulk equations of motion. These points can be systematically identified through a near-horizon analysis \cite{blake_2018_Upper_Chaos_pole_skipping_2, blake_2020_lower_infinite_pole_skipping_1} or through the covariant expansion formalism \cite{wang_2022_Pole-skipping_gauge_bosonicfields, ning_2023_pole-skipping_gauge_fermionicfields}, both of which act directly on the bulk equations of motion. To date, pole-skipping phenomena have been studied across a wide range of settings \cite{Haehl_2018_EFT_chaotic_CFT, Grozdanov_2019_higher_derivative_correction, grozdanov_2019_lower_infinite_pole_skipping_2, natsuume_2019_Gauss-Bonnet_correction_pole-skipping_2, wu_2019_gauss-Bonnet_correction_pole-skipping_k_change_w_unchange, ahn_2019_pole_skipping_Lower_chaos_hyperbolic, natsuume_2019_lower_infinite_pole_skipping_other, ceplak_2020_Pole_Skipping_fermion_BTZ_analytical, Liu_2020_TMG_chaos_pole-skipping, ahn_2020_Pole_skipping_chaos_hyperbolic_CFT, abbasi_2020_Pole_skipping_chaos_chiral_anomly, abbasi_2020_pole_skipping_Charged_fluid_2, liu_2020_pole_skipping_chaos_Topology, jansen_2020_pole_skipping_Charged_fluid_1, ahn_2021_Classify_pole_skipping, choi_2021_Pole_skipping_SYK_chain, jeong_2021_Pole_skipping_chaos_axion_model, ramirez_2021_Pole_skipping_chaos_CFT_2, natsuume_2021_Pole_skipping_zero_temperature, ceplak_2021_Pole_skipping_chaos_3/2_spin, blake_2021_pole-skipping_rotating_black_hole, yuan_2021_near-Horizon-analysis-Lif-Rindler-AdS, amano_2023_Pole_skipping_chaos_spinning_plasma, grozdanov_2023_Pole_skipping_hyperbolic_sphereical_flat, wang_2023_High_form_gauge_pole-skipping_deformation, Natsuume_2023_pole-skipping_non-black_hole, yuan_2023_Pole_skipping_SYK_Model, ahn_2024_Lower_pole-skipping_coupledEOM_LinearAxion, pan_2024_Pole_Skipping_stueckelberg, yuan_2024_Pole_skipping_SYK_double_scale, Lilani_2025_chaos_QCD}, revealing their ubiquity across diverse physical systems. This includes higher-derivative gravity theories \cite{Grozdanov_2019_higher_derivative_correction, natsuume_2019_Gauss-Bonnet_correction_pole-skipping_2, wu_2019_gauss-Bonnet_correction_pole-skipping_k_change_w_unchange, Liu_2020_TMG_chaos_pole-skipping} and purely field-theoretic models \cite{Haehl_2018_EFT_chaotic_CFT, ramirez_2021_Pole_skipping_chaos_CFT_2, choi_2021_Pole_skipping_SYK_chain}.

Previous studies have leveraged pole-skipping points to reconstruct the quasinormal mode spectra \cite{abbasi_2021_Pole_skipping_chaos_bound_QNM, grozdanov_2023_Pole_skipping_reconstruct_spectrum} and to establish bounds on transport coefficients \cite{grozdanov_2021_univalence_bound_pole-skipping, jeong_2021_Pole_skipping_chaos_axion_model, baggioli_2022_univalence_axion_models}. In particular, Ref. \cite{grozdanov_2023_Pole_skipping_reconstruct_spectrum} showed that the infinite tower of pole-skipping points aligned along a single hydrodynamic mode, expanded in large spacetime dimensions, is sufficient to reconstruct both the full quasinormal spectrum and the associated Green's function. This finding suggests that holographic Green's functions can be efficiently encoded with reduced information, with pole-skipping points serving as essential information carriers. 

While distinct in methodology and focus, our result resonates with the same principle: pole-skipping points encode essential information---in this case, the bulk metric itself. Specifically, we demonstrate that a discrete infinite set of lower-half pole-skipping points is sufficient to reconstruct the metric of a general static, planar-symmetric black hole in arbitrary dimension.  In ingoing Eddington-Finkelstein coordinates, such a spacetime is described by the metric: 

\begin{equation}\label{equ_general_metric_d} ds^2 = -g_{vv}(r) dv^2 + 2g_{vr}(r) dvdr + r^2 d\vec{x}^2, 
\end{equation} 
where $\vec{x}$ is a spatial vector of dimension $d$.\footnote{We choose the radial coordinate such that $g_{xx}(r) = r^{2}$, as discussed in Chapter 5 of \cite{Carroll_2004_spacetime_and_geometry}. This ansatz is widely used to describe AdS black holes.} The metric components $g_{vv}$ and $g_{vr}$ can be expanded around the horizon as:\footnote{In all subsequent discussions, we refer to the \textbf{horizon} as the outer one if the black hole has multiple horizons.} \begin{equation}\label{equ_gvvgvr_expansion}
\begin{aligned}
& g_{vv}(r)=g_{vv_1}(r-r_h)+g_{vv_2}(r-r_h)^2+\ldots \\
& g_{vr}(r)=g_{vr_0}+g_{vr_1}(r-r_h)+g_{vr_2}(r-r_h)^2+\ldots
\end{aligned}
\end{equation}
Since this metric represents a black hole with a horizon at $r=r_{h}$, we have set $g_{vv_{0}}=0$.

Our reconstruction method allows us to analytically reinterpret the near-horizon expansion coefficients $g_{vv_{n}}$ and $g_{vr_{n-1}}$ in terms of boundary pole-skipping data to arbitrary order, by iteratively solving sets of linear equations. The same procedure applies to the near-horizon expansion of the vacuum Einstein equations, enabling a corresponding reformulation of gravitational information in terms of pole-skipping data alone.

Notably, our reconstruction method also applies to black hole spacetimes with Lifshitz scaling and hyperscaling violation, and remains valid under $T\bar{T}$ deformations of the boundary QFT.

By carrying out the reconstruction to sufficiently high order $n$, one can recover the metric functions $g_{vv}(r)$ and $g_{vr}(r)$ with arbitrary precision, extending from the horizon to any desired region of the spacetime, including both the black hole interior and exterior, provided the near-horizon series remain holomorphic throughout the relevant domain in the analytically continued complex-$r$ plane. This is remarkable, as most previous methods reconstruct only the exterior, with complexity-based approaches \cite{hashimoto_2021_complexity_bulk_metric_1, xu_2023_Entanglement_reconstruct_bulk_metric_ML_1} being the sole alternative for accessing the interior.


Additionally, our reconstruction reveals a special algebraic structure underlying the pole-skipping points of certain Klein-Gordon equations of the form $(\nabla^{2} + V(r))\Phi = 0$. This structure manifests as an infinite tower of homogeneous polynomial identities in the pole-skipping momenta, which we refer to as $\mu$-polynomial constraints. These constraints are strong enough to determine the locations of most pole-skipping points purely algebraically. Importantly, they exhibit a striking universality: their validity does not depend on the specific details of the bulk geometry, such as its dimensionality, asymptotic structure, or the presence of a holographic dual.

This paper is organized as follows. In Section \ref{sec_PS_review}, we briefly revisit the procedure for identifying pole-skipping points via near-horizon analysis in the extremal BTZ background. Following this, in Section \ref{sec_general_reconstruct}, we introduce and implement our reconstruction method for general static, planar symmetric black holes in arbitrary spatial dimension $d$, coupled to a massless probe scalar field. Then, in Section \ref{sec_reconstruct_extend}, we extend this method beyond the probe limit and apply it to bulk theories whose boundary duals exhibit Lifshitz scaling, hyperscaling violation, or $T\bar{T}$ deformations. Section \ref{sec_Einstein_equation_PS} shows how this reconstruction method offers a novel reinterpretation of the vacuum Einstein equations in terms of pole-skipping data. Subsequently, in Section \ref{sec_mu_polynomial_constraint}, we reveal that most pole-skipping points are redundant and conform to an equal number of homogeneous polynomial identities, termed $\mu$-polynomial constraints. We further present a universal recursive algorithm for systematically deriving these identities. The validity of these identities is then verified in various holographic models in Section \ref{sec_example_verify_polynomial_constraint}. We conclude the paper with a discussion in Section \ref{sec_discussion}. A concise summary of the results presented in this paper can be found in the letter version \cite{Lu_2025_Shorter_paper}.

\section{Pole-skipping points review}\label{sec_PS_review}

We start with a planar symmetric asymptotically AdS background coupled to a probe scalar field $\phi$ with mass $m$. In the context of holography, this scalar field $\phi$ is dual to a boundary CFT operator $\mathcal{O}$ with scaling dimension $\Delta$, which is related to $m$ through the equation


\begin{equation}
\label{equ_scaling_to_mass}
\Delta (\Delta-d-1)=m^2 L^2.
\end{equation}

If we choose the standard (alternative) quantization, the scaling dimension $\Delta$ of $\mathcal{O}$ corresponds to the larger (smaller) root of equation \eqref{equ_scaling_to_mass}. For the remainder of our discussion, we set the radius of AdS $L$ to be unity. Under the probe limit, the equation of motion for $\phi$ is governed by the Klein-Gordon equation $(\nabla^{2}+m^2)\phi=0$. Upon transitioning to Fourier space with the introduction of $\phi = \varphi(r)e^{-i\omega v+ikx}$, $\varphi$ typically exhibits asymptotic behavior near the AdS boundary as $\varphi=\varphi_{+}(\omega,k) r^{\Delta-d-1}+\varphi_{-}(\omega,k) r^{-\Delta}+\ldots$ This allows us to uniquely determine the Fourier-transformed retarded Green’s function of the boundary scalar operator $\mathcal{O}$:

\begin{equation}
\label{equ_GR_standard}
\mathcal{G}^{\mathcal{O}}_{R}(\omega, k)\propto (2\Delta-d-1)\frac{\varphi_{-}(\omega, k)}{\varphi_{+}(\omega, k)}.
\end{equation}
However, this framework encounters a breakdown at specific frequencies and momenta, known as pole-skipping points. At these points, the ingoing solution at the horizon loses its uniqueness, leading to an indeterminate value for $\mathcal{G}^{\mathcal{O}}_{R}(\omega,k)$ at these points \cite{blake_2018_upper_Chaos_pole_skipping_1,grozdanov_2018_upper_Chaos_pole_skipping_0,blake_2018_Upper_Chaos_pole_skipping_2,blake_2020_lower_infinite_pole_skipping_1}. To illustrate this phenomenon, we briefly review the pole-skipping in the extremal BTZ black hole, as studied in \cite{blake_2020_lower_infinite_pole_skipping_1}.

\subsection{Pole-skipping points from near-horizon analysis}\label{subsec_PS_nearhorizon}

The extremal BTZ black hole coupled with a probe scalar field is characterized by the following background:

\begin{equation}\label{equ_BTZ_metric}
ds^2=-(r^2-r_{h}^2) dt^2+\frac{1}{(r^2-r_{h}^2)}dr^2+r^2 dx^2,
\end{equation}
where $r_{h}$ denotes the location of the horizon. The related Hawking temperature is given by $T_{h}=\frac{r_{h}}{2\pi}$. For subsequent analysis, it is convenient to introduce the ingoing Eddington-Finkelstein (EF) coordinate $v$ defined as 

\begin{equation}\label{equ_ingoing_EF_Transform_BTZ}
\begin{aligned}v=t+r_*, \quad \frac{dr_*}{dr}=\frac 1{(r^2-r_{h}^2)},\end{aligned}
\end{equation}
where the metric \eqref{equ_BTZ_metric} becomes 

\begin{equation}\label{equ_ingoingEF_coordinate_BTZ}
\begin{aligned}ds^2&=-(r^2-r_{h}^2) dv^2+2 dvdr+r^{2}dx^2.\end{aligned}
\end{equation}
After performing the Fourier transformation $\phi = \varphi(r)e^{-i\omega v+ikx}$, the Klein-Gordon equation $(\nabla^{2}+m^2)\phi=0$ in the coordinates \eqref{equ_ingoingEF_coordinate_BTZ} can be explicitly written as

\begin{equation}\label{equ_KGequation_EFcoordinate}
-\frac{\left (\mu+m^2 r^2+i r \omega \right)}{r^2}\varphi (r)+\left (3 r-\frac{r_h^2}{r}-2 i \omega \right)\varphi' (r)+\left (r^2-r_h^2\right) \varphi'' (r)=0,
\end{equation}
where we define $\mu=k^2$ for convenience and will use it in most subsequent contexts. We then expand $\varphi(r)$ around the horizon with the ansatz

\begin{equation}\label{equ_Taylor_Expansion_rh}
\varphi (r)=(r-r_{h})^{\alpha}\sum_{p=0}^\infty\varphi_n (r-r_h)^p.
\end{equation}
There are two independent powers $\alpha$ for the expansion around the horizon in equation \eqref{equ_KGequation_EFcoordinate}:

\begin{equation}\label{equ_two_indices}
\alpha_1=0,\quad\alpha_2=\frac{i\omega}{2\pi T_{h}},
\end{equation}
where we keep $T_{h}$ unspecified. In ingoing EF coordinates \eqref{equ_ingoingEF_coordinate_BTZ}, the ingoing boundary condition near the horizon reduces to the regularity condition, which selects $\alpha_{1}=0$ among the two powers. This unique power choice ensures the uniqueness of the holographic Green’s function at the boundary.

However, at the pole-skipping frequencies: $\omega_{n}=-i2\pi n T_{h}$ with $n\in\mathbb{Z}^{+}$, both $\alpha_{1}=0$ and $\alpha_{2}=n$ appear to yield regular near-horizon solutions, seemingly violating the uniqueness of the Green’s function. In reality, a unique ingoing solution still exists, as the solution associated with $\alpha_{2}$ develops logarithmic corrections $\log(r - r_h)$ that spoil regularity at the horizon. Yet, as highlighted in \cite{blake_2020_lower_infinite_pole_skipping_1}, these logarithmic corrections vanish when the momentum squared $\mu$ assumes pole-skipping values $\mu_{n}$. At these pole-skipping points $(\omega_{n},\mu_{n})$, one can express the solution $\varphi (r)$ as a combination of two ingoing solutions:

\begin{equation}
\label{equ_Two_ingoing_Sol}
\varphi (r)=\varphi_0+\varphi_{1}(r-r_{h})+\ldots+\varphi_n (r-r_{h})^n+\varphi_{n+1}(r-r_{h})^{n+1}+\ldots,
\end{equation}
where $\varphi_0$ and $\varphi_n$ are independent parameters, while remaining coefficients are determined accordingly. The presence of two free parameters near the horizon indicates that the ingoing solution is no longer unique at $(\omega_n, \mu_n)$, rendering the boundary Green’s function $\mathcal{G}^{\mathcal{O}}_R(\omega, \mu)$ ambiguous and ill-defined at these points.

Those pole-skipping points can be determined by a near-horizon analysis developed in \cite{blake_2020_lower_infinite_pole_skipping_1}. To initiate the analysis, we substitute expansion \eqref{equ_Taylor_Expansion_rh} into the equation of motion \eqref{equ_KGequation_EFcoordinate} and impose the condition that each coefficient in front of $(r-r_{h})^{n-1}$ is zero. This procedure yields a set of $n$ linear equations of the form:

\begin{equation}\label{equ_Linear_equation_horizonexpansion}
\mathbb{M}\cdot\varphi=\begin{pmatrix}M_{11}&2 (2\pi T_{h}-i\omega)&0&0&\cdots&0\\M_{21}&M_{22}&4 (4\pi T_{h}-i\omega)&0&\cdots&0\\M_{31}&M_{32}&M_{33}&6 (6\pi T_{h}-i\omega)&\cdots&0\\\vdots&\vdots&\vdots&\vdots&\ddots&\vdots\\M_{n 1}&M_{n 2}&M_{n 3}&M_{n 4}&\cdots&2 n (2 n\pi T_{h}-i\omega)\end{pmatrix}\begin{pmatrix}\varphi_0\\\varphi_1\\\varphi_2\\\vdots\\\varphi_{n}\end{pmatrix}=\begin{pmatrix}0\\0\\0\\\vdots\end{pmatrix}.
\end{equation}
The parameter matrix $\mathbb{M}$ has dimensions of $n\times (n+1)$, where $M_{ij}$ are functions in terms of $\omega$ and $\mu$. The dimension of the solution space, spanned by the elements $\{\varphi_{0}, \varphi_{1}, \dots, \varphi_{n}\}$, denoted as $N$, equals the total number of variables, $n+1$, minus the rank of $\mathbb{M}$, expressed as $N = n + 1 - \text{Rank}(\mathbb{M})$. For general values of $\omega$ and $k$, the rank of $\mathbb{M}$ is $n$, indicating the count of its linearly independent columns. Thus, only one parameter, $\varphi_{0}$, is independent. However, the situation changes at pole-skipping points $(\omega_{n}, \mu_{n})$. Here, the determinant of the first $n \times n$ sub-matrix $\mathcal{M}^{(n)}$ of $\mathbb{M}$ becomes zero, and the $(n+1)^\text{th}$ column vanishes. These conditions reduce the rank of $\mathbb{M}$ to $n-1$, resulting in $N = 2$, indicating two independent parameters, $\varphi_{0}$ and $\varphi_{n}$. This aligns with our earlier assertion that two ingoing solutions exist at pole-skipping points. 

The above demonstration also provides insight into the determination of the pole-skipping points $(\omega_{n}, \mu_{n})$. The value of $\omega_{n}$ can be obtained by setting the last column of $\mathbb{M}$ to zero, expressed as $2n(2n\pi T_{h}-i\omega)=0$, resulting in $\omega_{n}=-i2 n\pi T_{h}$. To locate the positions of $\mu_{n}$, we substitute the obtained $\omega_{n}$ into $\mathcal{M}^{(n)}$ and search for $\mu=\mu_{n}$ such that the determinant of $\mathcal{M}^{(n)}(\omega_{n},\mu_{n})$ becomes zero. In other words, we solve the equation $\text{Det}(\mathcal{M}^{(n)}(\omega_n, \mu)) = 0$, or more succinctly, $\text{Det}(\mathcal{M}^{(n)}(\boldsymbol{\mu})) = 0$ for $\mu$.

To illustrate the preceding discussion, we collect the linear equations up to the second order and organize them in matrix form as follows:
\begin{equation}\label{equ_First_two_KGexpansion}
\begin{pmatrix}M_{11}&2 (2\pi T_{h}-i\omega)&0\\M_{21}&M_{22}&4 (4\pi T_{h}-i\omega)\end{pmatrix}\begin{pmatrix}\varphi_0\\\varphi_1\\\varphi_2\end{pmatrix}=\begin{pmatrix}0\\0\end{pmatrix}.
\end{equation}
In the BTZ case, the explicit expressions for $M_{11}$, $M_{21}$, and $M_{22}$ are given by:

\begin{equation}\label{equ_M_00_M_10_M_11}
M_{11}=-\frac{\mu+m^2 r_h^2+ir_h\omega}{r_h^2},~M_{21}=\frac{2 \mu+ir_h\omega}{r_h^3},~M_{22}=-\frac{\mu+r_h (r_h (m^2-4)+i\omega)}{r_h^2}.
\end{equation}
To determine the pole-skipping points $(\omega_2,\mu_2)$, we can first obtain $\omega_{2}=-i 4\pi T_{h}$ by solving $4(4\pi T_{h}-i\omega)=0$. To locate $\mu_{2}$, we solve $\text{Det}(\mathcal{M}^{(2)}(\boldsymbol{\mu}))=0$, yielding 

\begin{equation}\label{equ_second_pole-skipping_BTZ}
\begin{aligned}\omega_{2}=-2 i r_{h},\quad\quad \mu_2=-r_h^2 (\Delta-2)^2,-r_h^2\Delta^2.
\end{aligned}
\end{equation}
where we substitute $T_{h}=\frac{r_{h}}{2\pi}$ and express the mass $m$ of $\varphi$ in terms of the scaling dimension $\Delta$ of $\mathcal{O}$ using $m^{2}=\Delta(\Delta-2)$.

The same method can be applied to higher-dimensional matrices $\mathbb{M}$, which include all the linear equations up to higher orders $n$. In the BTZ case, as shown in \cite{blake_2020_lower_infinite_pole_skipping_1}, the determinant $\text{Det}(\mathcal{M}^{(n)}(\boldsymbol{\mu})) = 0$ is a polynomial of order $n$ with respect to $\mu$, implying $n$ solutions. Alongside $\omega_{n}=-i2\pi n T_{h}$, the general form of pole-skipping points can be expressed as:

\begin{equation} \label{equ_BTZ_pole-skipping_allorder} \omega_{n}=-i r_{h} n,\quad \mu_{n, q}=-r_{h}^{2}(n-2 q+\Delta)^2, \end{equation}
where $q$ belongs to the set $\{1,\ldots,n\}$.

By definition, the retarded Green’s function $\mathcal{G}^{\mathcal{O}}_{R}(\omega,k)$ in the boundary CFT becomes ill-defined at these points. We directly verify this in the next subsection. 

\subsection{Green's functions behavior at pole-skipping points}\label{subsec_QFTGF_PS}

For the BTZ metric \eqref{equ_BTZ_metric}, the retarded Green's function for the scalar operator in the boundary CFT can be analytically obtained as:

\begin{equation}\label{equ_BTZ_GR} 
\mathcal{G}^{\mathcal{O}}_{R}(\omega, k)\propto\frac{\Gamma\left (\frac{\Delta}{2}+\frac{i (k-\omega)}{4\pi T_{b}}\right)\Gamma\left (\frac{\Delta}{2}-\frac{i (k+\omega)}{4\pi T_{b}}\right)}{\Gamma\left (1-\frac{\Delta}{2}+\frac{i (k-\omega)}{4\pi T_{b}}\right)\Gamma\left (1-\frac{\Delta}{2}-\frac{i (k+\omega)}{4\pi T_{b}}\right)}, 
\end{equation}
where $T_b$ represents the boundary temperature. The Gamma function has no zeros and only single poles at non-positive integers. Therefore, the poles of the Gamma function in the numerator contribute to the poles of $\mathcal{G}^{\mathcal{O}}_{R}(\omega,k)$, which are explicitly given by:

\begin{equation}\label{equ_poles_BTZ_GR} \omega_{p^\pm}^n(k)=\pm k-i 2\pi T_b (\Delta+2 n), \end{equation}
while the poles of the Gamma function in the denominator give rise to the zeros of $\mathcal{G}^{\mathcal{O}}_{R}(\omega,k)$: 

\begin{equation}\label{equ_zeroes_BTZ_GR}
    \omega_{z^\pm}^{n}(k)=\pm k-i2\pi T_b(2-\Delta+2n),
\end{equation}
where $n$ belongs to $\{0,1,2,\dots\}$.

\begin{figure}[h!]
\centering
 \includegraphics[width=1\textwidth]{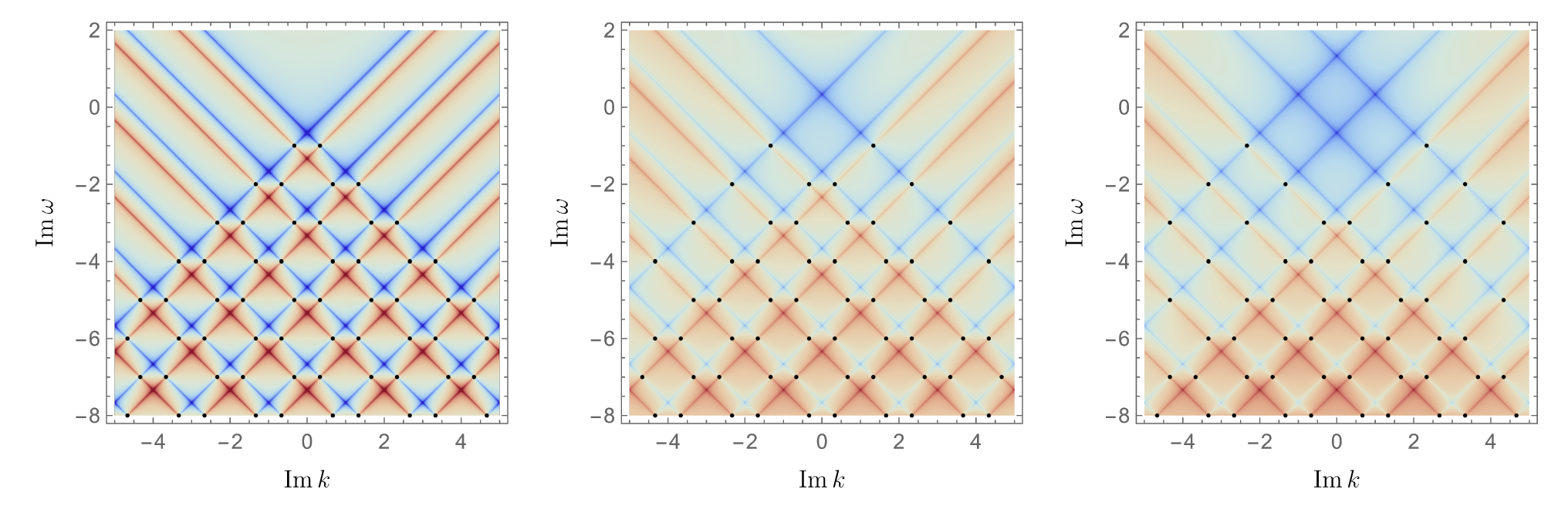}
\caption{\label{fig_BTZ_Green_density}From left to right, the figure displays heatmaps showing the values of $\log|\mathcal{G}^{\mathcal{O}}_{R}(\mathrm{Im}\omega,\mathrm{Im}k)|$ for different values of $\Delta$: $\frac{4}{3}$, $\frac{7}{3}$, and $\frac{10}{3}$ (after taking $T_{b}=\frac{1}{2\pi}$). The red lines and blue lines represent the poles and zeroes of Eq. \eqref{equ_BTZ_GR} respectively. The black dots depict the pole-skipping points obtained from the near-horizon analysis. In all panels, the locations of the black dots precisely coincide with the intersections of poles and zeroes of $\mathcal{G}^{\mathcal{O}}_{R}(\omega,k)$.}
\end{figure}

Consider now the scenario where both the numerator and denominator Gamma functions have poles, resulting in the values of $\mathcal{G}^{\mathcal{O}}_{R}(\omega,k)$ taking the form $\frac{\infty}{\infty}$ and thus becoming indeterminate. Alternatively, this situation can be interpreted as the intersections between the lines of poles and lines of zeroes of the $\mathcal{G}^{\mathcal{O}}_{R}(\omega,k)$, which can be determined by solving $\omega_{p^\pm}^{i}(k)=\omega_{z^\pm}^{j}(k)$. This equation yields the following solution:

\begin{equation}\label{equ_intersections_poles_zeroes_BTZ} \omega_{m}=-i 2\pi T_b m,\quad \mu_{m,q}=-4\pi^2 T_{b}^2 (m-2 q+\Delta)^2, \end{equation}
where $m$ starts from $1$ and $q\in\{1,\ldots,m\}$. Upon identifying $T_b$ with the Hawking temperature $T_h = \frac{r_h}{2\pi}$, these locations precisely match those predicted from the near-horizon analysis \eqref{equ_BTZ_pole-skipping_allorder}. This alignment can be intuitively illustrated in Figure \ref{fig_BTZ_Green_density}.

We conclude this section with a discussion on the case of integral $\Delta$. Note that the expression \eqref{equ_BTZ_GR} is applicable only for non-integral values of $\Delta$. When $\Delta$ is an integer, the form of $\mathcal{G}^{\mathcal{O}}_{R}(\omega,k)$ differs, yet its pole-skipping points can still be characterized by the formula \eqref{equ_BTZ_pole-skipping_allorder}. The significant difference arises from specific pole-skipping points that emerge due to intersections between multiple curves of poles and multiple curves of zeros \cite{blake_2020_lower_infinite_pole_skipping_1, ahn_2021_Classify_pole_skipping}. Since the values of $\mathcal{G}^{\mathcal{O}}_{R}(\omega,k)$ remain indeterminate at these points, they are also categorized as pole-skipping points as per the definition and do not interfere with our reconstruction. They are termed as anomalous points in \cite{blake_2020_lower_infinite_pole_skipping_1} and type II pole-skipping points in \cite{ahn_2021_Classify_pole_skipping}.

\section{Reconstructing general metrics via pole-skipping points}\label{sec_general_reconstruct}

In the last section, we reviewed the definition of pole-skipping points and how their locations can be determined via near-horizon analysis for a given bulk metric. We now reverse the question: given access to some or all of the pole-skipping locations, can we reconstruct the bulk metric itself? The answer is affirmative. In what follows, we present a step-by-step procedure for carrying out this reconstruction.


We present our reconstruction method in the context of general static, planar-symmetric black holes expressed in ingoing EF coordinates, where the metric takes the form \eqref{equ_general_metric_d}. The metric components $g_{vv}(r)$ and $g_{vr}(r)$ admit Taylor expansions near the horizon at $r = r_{h}$, as given in Eq. \eqref{equ_gvvgvr_expansion}. Unless otherwise stated, we will set $r_{h}=1$ throughout the reconstruction, without loss of generality. In this setting, the Hawking temperature is given by $T_{h} = \frac{g_{vv_{1}}}{4\pi g_{vr_{0}}}$. Then, we introduce a massless probe scalar field $\phi$ satisfying the Klein–Gordon equation $\nabla^{2}\phi = 0$. Within the holographic framework, this corresponds to a thermal boundary CFT at temperature $T_{b}$, where $\phi$ is dual to a marginal operator $\mathcal{O}$ with scaling dimension $\Delta = d + 1$.

In the context of the general metric \eqref{equ_general_metric_d}, the Klein-Gordon equation for the Fourier mode $\varphi$ associated with the probe scalar field $\phi$ can be explicitly expressed as:

\begin{equation}\label{equ_KGequation_General_d_mass_zero}
\begin{aligned}
&\frac{g_{vv}(r)}{g_{vr}(r)^{2}}\varphi^{\prime\prime}(r)+\left (-\frac{2 i\omega}{g_{vr}(r)} + \frac{d g_{vv}(r)}{rg_{vr}(r)^{2}} - \frac{g_{vv}(r) g_{vr}^{\prime}(r)}{g_{vr}(r)^{3}} + \frac{g_{vv}^{\prime}(r)}{g_{vr}(r)^{2}}\right)\varphi^{\prime}(r)\\
&- \left (\frac{\mu}{r^{2}}+\frac{i d\omega}{r g_{vr}(r)}\right)\varphi (r)=0.\\
\end{aligned}
\end{equation}

By expanding the above equation to the $n^{\text{th}}$ order, one obtains an $n \times (n+1)$ parameter matrix $\mathbb{M}$, analogous to Eq. \eqref{equ_Linear_equation_horizonexpansion}, with the only non-vanishing entry in the last column given by $\mathbb{M}_{n,n+1}=\frac{n (n \pi T_{h}-i \omega )}{g_{vr_{0}}}$. Following the analysis in Section \ref{subsec_PS_nearhorizon}, the rank of $\mathbb{M}$ decreases from $n$ to $n-1$ when both of the following conditions are satisfied: the last column vanishes, yielding $\omega_{n} = -i 2 n \pi T_{h}$, and the determinant of the upper-left $n \times n$ submatrix $\mathcal{M}^{(n)}$ of $\mathbb{M}$ vanishes, leading to the condition $\text{Det}(\mathcal{M}^{(n)}(\boldsymbol{\mu})) = 0$ with $\omega$ fixed to $\omega_{n}$. This reasoning readily extends to equations of the form $(\nabla^{2}+V(r))\phi(r)=0$, where $\nabla$ is defined on the metric \eqref{equ_general_metric_d}. In particular, the determinant equation $\text{Det}(\mathcal{M}^{(n)}(\boldsymbol{\mu}))=0$, derived from Eq. \eqref{equ_KGequation_General_d_mass_zero} (and likewise from Eqs. \eqref{equ_KGequation_General_d} and \eqref{equ_KGequation_hyperviolation_d} introduced later), can be shown to yield a degree-$n$ polynomial in $\mu$, which can be written explicitly as

\begin{equation}\label{equ_mu_order_n_vieta}
V_{n, n}\mu^{n}+V_{n, n-1}\mu^{n-1}+\cdots+V_{n, 1}\mu+V_{n, 0}=0,
\end{equation}
where each $V_{n,m}$ is the coefficient of the term $\mu^{m}$ in $\text{Det}(\mathcal{M}^{(n)}(\boldsymbol{\mu}))$. According to the fundamental theorem of algebra, this polynomial admits $n$ roots, which can be denoted as $\mu_{n,q}$ with $q = 1, \ldots, n$. Due to the inherent $S_n$ symmetry within Eq. \eqref{equ_mu_order_n_vieta}, one can construct $n$ elementary symmetric polynomials $E_n(\mu^m)$ in the roots, where $m = 1, \ldots, n$ denotes the degree \cite{macdonald_1999_symmetric_polynomial}. For instance, $E_3(\mu^2) = \mu_{3,1}\mu_{3,2} + \mu_{3,2}\mu_{3,3} + \mu_{3,3}\mu_{3,1}$.

\subsection{Flipping the near-horizon analysis}\label{subsec_flip_near-horizon_analysis}

Unlike the near-horizon analysis in Section \ref{sec_PS_review}, which aims to determine the locations of pole-skipping points, our reconstruction procedure assumes the pole-skipping points $(\omega_{n}, \mu_{n,q})$ are already known.\footnote{The condition $\omega_n = -i 2\pi n T$ has already been established in the analysis below Eq. \eqref{equ_KGequation_General_d_mass_zero}.} These points mark the locations where the boundary Green's function $\mathcal{G}^{\mathcal{O}}_{R}(\omega,k)$ becomes ambiguous. Instead, the unknown variables are the near-horizon expansion coefficients $g_{vv_{n}}$ and $g_{vr_{n-1}}$ in \eqref{equ_gvvgvr_expansion}. This reversal of roles transforms the original determinant equation $\text{Det}(\mathcal{M}^{(n)}(\boldsymbol{\mu})) = 0$, derived from the $n^{\text{th}}$-order near-horizon analysis, into its `flipped' counterpart $\text{Det}(\mathcal{M}^{(n)}(g_{vv_{n}}, g_{vr_{n-1}})) = 0$, or more succinctly, $\text{Det}(\vec{\mathcal{M}}^{(n)}(\mathbf{g})) = 0$. This yields a system of $n$ equations in terms of $g_{vv_{n}}$ and $g_{vr_{n-1}}$, with each equation corresponding to a distinct choice of $\mu_{n,q}$.

Yet a more natural and insightful formulation arises by exploiting the $S_n$ symmetry and labeling the equations in $\text{Det}(\vec{\mathcal{M}}^{(n)}(\mathbf{g})) = 0$ using the $n$ elementary symmetric polynomials $E_n(\mu^m)$. These polynomials are connected to the coefficients $g_{vv_{i}}$ and $g_{vr_{i-1}}$ through the polynomial coefficients $V_{n,m}$ introduced in Eq. \eqref{equ_mu_order_n_vieta}. Specifically, by applying Vieta's formula, we obtain
\begin{equation}\label{equ_elementary_symmetric_polynomial_vieta}
E_{n}(\mu^{m})=\frac{v_{n, n-m}}{v_{n, n}},
\end{equation}
where, for convenience, we define $v_{n,m} = (-1)^{n-m} V_{n,m}$. In this formulation, the quantities $E_n(\mu^m)$ are known inputs, determined from the given set of pole-skipping points, while the $v_{n,m}$ are generally algebraic functions of unknowns: $g_{vv_{i}}$ and $g_{vr_{i-1}}$. The $q^{\text{th}}$ equation in $\text{Det}(\vec{\mathcal{M}}^{(n)}(\textbf{g}))=0$, denoted as $\text{Det}(\mathcal{M}^{(n)}_q(\mathbf{g})) = 0$, is thus equivalent to the identity

\begin{equation}\label{equ_qth_equation}
E_n(\mu^q) - \frac{v_{n, n-q}}{v_{n, n}} = 0.
\end{equation}

The distinction between $\text{Det}(\mathcal{M}^{(n)}(\boldsymbol{\mu})) =0$ and $\text{Det}(\vec{\mathcal{M}}^{(n)}(\textbf{g}))=0$ is summarized in Table \ref{tab_compare_two_Det}.
\begin{table}[h]
    \centering
    \begin{tabular}{|c|c|c|}
        \hline
         & $\text{Det}(\mathcal{M}^{(n)}(\boldsymbol{\mu})) =0$ & $\text{Det}(\vec{\mathcal{M}}^{(n)}(\textbf{g}))=0$ \\
        \hline
        Fixed parameters & $g_{vv_{n}},g_{vr_{n-1}}$ & $\mu_{n,q}$ \\
        \hline
        Variables to solve & $\mu_{n,q}$ & $g_{vv_{n}},g_{vr_{n-1}}$ \\
        \hline
        Representation & \textbf{One} polynomial equation of degree $n$ & \textbf{n} algebraic equations \\
        \hline
    \end{tabular}
    \caption{Comparison between $\text{Det}(\mathcal{M}^{(n)}(\boldsymbol{\mu})) =0$ and $\text{Det}(\vec{\mathcal{M}}^{(n)}(\textbf{g}))=0$}
   \label{tab_compare_two_Det}
\end{table}

\subsection{Reconstruction of $g_{vv_{n}}$ and $g_{vr_{n-1}}$ for $n=1,2$}\label{subsec_fix_Delta&M_relation}

We start our reconstruction procedure by expanding Eq. \eqref{equ_KGequation_General_d_mass_zero} at the horizon up to the first order ($n=1$). This expansion yields a single equation $\text{Det}(\vec{\mathcal{M}}^{(1)}(\textbf{g}))=0$ (equivalently $\text{Det}(\mathcal{M}^{(1)}_{1}(\textbf{g}))=0$) with the form:

\begin{equation}\label{equ_M11_general}
    E_{1}(\mu)+\frac{d g_{vv_{1}}}{2 g_{vr_{0}}^{2}}=0.
\end{equation}
To close the system, we align the boundary temperature $T_b$, expressed as $T_b = i\frac{\omega_1}{2\pi}$, with the Hawking temperature $T_h = \frac{g_{vv_1}}{4\pi g_{vr_0}}$. This yields the second equation:

\begin{equation}\label{equ_Tb_Th}
    i\frac{\omega_1}{2\pi} = \frac{g_{vv_1}}{4\pi g_{vr_0}}.
\end{equation}

Combining Eq. \eqref{equ_M11_general} and Eq. \eqref{equ_Tb_Th} yields explicit expressions for $g_{vv_{1}}$ and $g_{vr_{0}}$ in terms of $\omega_{1}$ and $E_{1}(\mu)$ (i.e., $\mu_{1,1}$):

\begin{equation}\label{equ_sol_gvv1_gvr0_d}
g_{vv_1}=\frac{2 d \omega_1^2}{E_{1}(\mu)}, \quad g_{vr_0}=-\frac{i d \omega_1}{E_{1}(\mu)}.
\end{equation}

We proceed by examining the case $n=2$, i.e., expanding the Klein–Gordon equation \eqref{equ_KGequation_General_d_mass_zero} to second order. Substituting the first-order expressions for $g_{vv_{1}}$ and $g_{vr_{0}}$ from Eq. \eqref{equ_sol_gvv1_gvr0_d}, the equations $\text{Det}(\mathcal{M}^{(2)}_{1}(\mathbf{g})) = 0$ and $\text{Det}(\mathcal{M}^{(2)}_{2}(\mathbf{g})) = 0$, corresponding respectively to $E_2(\mu) - \frac{v_{2,1}}{v_{2,2}} = 0$ and $E_2(\mu^2) - \frac{v_{2,0}}{v_{2,2}} = 0$, take the form:

\begin{equation}\label{equ_Det_2_equ}   
\begin{aligned}
&E_{2}(\mu)-\frac{4 E_{1}(\mu)}{d}-2 E_{1}(\mu)+\frac{2 E_{1}(\mu)^{2} g_{vv_2}}{d^2 \omega_1^2}-\frac{2 i E_{1}(\mu)^{2}g_{vr_1}}{d^2 \omega_1}= 0,\\
&E_{2}(\mu^2)-\frac{4 E_{1}(\mu)^2}{d} + \frac{4 E_{1}(\mu)^3 g_{vv_2}}{d^2 \omega_1^2} - \frac{8 i E_{1}(\mu)^3 g_{vr_1}}{d^2 \omega_1} = 0.\\
\end{aligned}
\end{equation}
The set of equations presented above comprises two linear equations in terms of $g_{vv_{2}}$ and $g_{vr_{1}}$. It provides the solutions for $g_{vv_{2}}$ and $g_{vr_{1}}$ as  

\begin{equation}\label{equ_sol_gvv2_gvr1_d}
    \begin{aligned}
       &g_{vv_{2}}=\frac{d^2 \omega_1^2 E_{2}(\mu^2)}{4 E_{1}(\mu)^3} + \frac{2 d^2 \omega_1^2}{E_{1}(\mu)} - \frac{d^2 \omega_1^2 E_{2}(\mu)}{E_{1}(\mu)^2} + \frac{3 d \omega_1^2}{E_{1}(\mu)},\\ 
       &g_{vr_{1}}=\frac{i d^2 \omega_1 E_{2}(\mu)}{2 E_{1}(\mu)^2} - \frac{i d^2 \omega_1}{E_{1}(\mu)} - \frac{i d^2 \omega_1 E_{2}(\mu^2)}{4 E_{1}(\mu)^3} - \frac{i d \omega_1}{E_{1}(\mu)}.\\
    \end{aligned}
\end{equation}

\subsection{Reconstruction of $g_{vv_{n}}$ and $g_{vr_{n-1}}$ for $n\geq3$}\label{subsec_introduce_linear_mu_constraints}

For larger $n$, the system $\text{Det}(\vec{\mathcal{M}}^{(n)}(\mathbf{g})) = 0$ becomes overdetermined, involving two unknowns: $g_{vv_{n}}$ and $g_{vr_{n-1}}$, but comprising $n$ equations. Nevertheless, this does not obstruct the reconstruction. As proven in Appendix \ref{App_linearity_prove}, once the previously determined coefficients $g_{vv_{m}}$ and $g_{vr_{m-1}}$ for $m < n$ are substituted into $\text{Det}(\vec{\mathcal{M}}^{(n)}(\mathbf{g})) = 0$, only the final two equations---namely, $\text{Det}(\mathcal{M}^{(n)}_{n-1}(\mathbf{g})) = 0$ and $\text{Det}(\mathcal{M}^{(n)}_{n}(\mathbf{g})) = 0$---remain \textbf{linear} dependent on $g_{vv_{n}}$ and $g_{vr_{n-1}}$. Therefore,  for $n > 2$, the procedure to solve for $g_{vv_{n}}$ and $g_{vr_{n-1}}$ mirrors that for $n = 2$: 

\begin{enumerate}
\item Incorporate all previously obtained solutions from $\text{Det}(\vec{\mathcal{M}}^{(m)}(\mathbf{g})) = 0$ with $m < n$ into $\text{Det}(\vec{\mathcal{M}}^{(n)}(\textbf{g}))=0$, yielding two linear equations for $g_{vv_{n}}$ and $g_{vr_{n-1}}$: $\text{Det}(\mathcal{M}^{(n)}_{n-1}(\mathbf{g})) = 0$ and $\text{Det}(\mathcal{M}^{(n)}_{n}(\mathbf{g})) = 0$.

\item Solve these two equations, or equivalently, solve

\begin{equation}\label{equ_last_two_equations_vieta}
E_n (\mu^{n-1})- \frac{v_{n, 1}}{v_{n, n}}=0, \quad E_n(\mu^n)- \frac{v_{n, 0}}{v_{n, n}}=0,
\end{equation}
to determine $g_{vv_{n}}$ and $g_{vr_{n-1}}$.
\end{enumerate}


Note that the two equations in Eq. \eqref{equ_last_two_equations_vieta} are linearly independent, as each involves an elementary symmetric polynomial of a different degree. Provided a solution exists, it follows that for all $n \geq 2$, the pair $\text{Det}(\mathcal{M}^{(n)}_{n-1}(\mathbf{g})) = 0$ and $\text{Det}(\mathcal{M}^{(n)}_{n}(\mathbf{g})) = 0$ uniquely determines $g_{vv_n}$ and $g_{vr_{n-1}}$. Both variables can be fully expressed in terms of the pole-skipping data: $\omega_1$, $E_m(\mu^{m-1})$, and $E_m(\mu^{m})$ for all $m \leq n$. In general, the expressions for $g_{vv_{n}}$ and $g_{vr_{n-1}}$ become increasingly lengthy for $n > 2$ and are impractical to display. As a representative example, we provide the explicit forms of $g_{vv_{3}}$ and $g_{vr_{2}}$ as 
\begin{equation}\label{equ_sol_gvv3_gvr2_d}
\hspace*{-0.75cm}
    \begin{aligned}
&g_{vv_{3}}=\frac{d^3 \omega_1^2 E_{2}(\mu^2)^2}{16 E_{1}(\mu)^5} + \frac{d^3 \omega_1^2 E_{3}(\mu^2)}{8 E_{1}(\mu)^3} + \frac{13 d^3 \omega_1^2}{12 E_{1}(\mu)} - \frac{d^3 \omega_1^2 E_{2}(\mu)}{2 E_{1}(\mu)^2} - \frac{d^3 \omega_1^2 E_{2}(\mu) E_{2}(\mu^2)}{4 E_{1}(\mu)^4}\\
&- \frac{d^3 \omega_1^2 E_{3}(\mu^3)}{72 E_{1}(\mu)^4}+\frac{d^2 \omega_1^2 E_{2}(\mu^2)}{4 E_{1}(\mu)^3} + \frac{2 d^2 \omega_1^2}{E_{1}(\mu)} - \frac{d^2 \omega_1^2 E_{2}(\mu)}{E_{1}(\mu)^2} + \frac{2 d \omega_1^2}{3 E_{1}(\mu)},\\
&g_{vr_{2}}=\frac{i d^3 \omega_1 E_{2}(\mu)}{4 E_{1}(\mu)^2} + \frac{3 i d^3 \omega_1 E_{2}(\mu) E_{2}(\mu^2)}{16 E_{1}(\mu)^4} + \frac{i d^3 \omega_1 E_{3}(\mu^3)}{48 E_{1}(\mu)^4} - \frac{i d^3 \omega_1}{2 E_{1}(\mu)} - \frac{i d^3 \omega_1 E_{2}(\mu^2)}{8 E_{1}(\mu)^3}\\
&- \frac{i d^3 \omega_1 E_{3}(\mu^2)}{16 E_{1}(\mu)^3} - \frac{3 i d^3 \omega_1 E_{2}(\mu^2)^2}{32 E_{1}(\mu)^5}+ \frac{i d^2 \omega_1 E_{2}(\mu)}{4 E_{1}(\mu)^2} - \frac{i d^2 \omega_1}{2 E_{1}(\mu)} - \frac{i d^2 \omega_1 E_{2}(\mu^2)}{8 E_{1}(\mu)^3}.\\
    \end{aligned}
\end{equation}

As a quick validation of our reconstruction method, we apply it to the BTZ metric \eqref{equ_BTZ_metric}. Using the BTZ pole-skipping data \eqref{equ_intersections_poles_zeroes_BTZ} with $T_{b} = \frac{1}{2\pi}$, $\omega_{1}=-i$ and $\Delta = 2$, we compute the first few values of $E_{n}(\mu^{m})$ as inputs for the reconstruction:
\begin{equation}\label{equ_BTZ_Enmum_3}  
E_{1}(\mu) = -1, \quad E_{2}(\mu) = -4, \quad E_{2}(\mu^{2}) = 0, \quad E_{3}(\mu^{2}) = 19, \quad E_{3}(\mu^{3}) = -9.  
\end{equation}
Substituting these into Eqs. \eqref{equ_sol_gvv1_gvr0_d}, \eqref{equ_sol_gvv2_gvr1_d}, and \eqref{equ_sol_gvv3_gvr2_d}, we recover the first few near-horizon expansion coefficients:

\begin{equation}\label{equ_sol_gvvn_gvrn-1_up_to_3_BTZ}  
\begin{aligned}  
&g_{vv_{1}} = 2, \quad g_{vr_{0}} = 1,\\  
&g_{vv_{2}} = 1, \quad g_{vr_{1}} = 0,\\  
&g_{vv_{3}} = 0, \quad g_{vr_{2}} = 0.\\ 
\end{aligned}  
\end{equation}
These coefficients exactly match the first three orders of the near-horizon expansion of the BTZ metric \eqref{equ_ingoingEF_coordinate_BTZ}.

By reconstructing the near-horizon expansion coefficients $g_{vv_{n}}$ and $g_{vr_{n-1}}$ to sufficiently large $n$, we can approximate $g_{vv}(r)$ and $g_{vr}(r)$ within the convergence disk of the near-horizon series to arbitrary precision. The radius of convergence is determined by the distance from the horizon at $r = 1$ to the nearest singularities in the analytically continued complex $r$-plane. The interior geometry can thus be reconstructed, assuming no singularities appear before reaching the central singularity at $r = 0$. Likewise, by introducing the coordinate transformation $z = \frac{1}{r}$, where the boundary lies at $z = 0$, the exterior geometry can also be reconstructed, as long as no singularities are encountered within the domain. For well-studied holographic black holes such as Schwarzschild–AdS and Reissner–Nordström–AdS, the only singularities in the metric components lie at the black hole center and at the spacetime boundary. In such cases, our method faithfully reconstructs the entire geometry, both inside and outside the horizon, under the coordinate choice described above. In more general scenarios, additional singularities may appear in the complex plane. For instance, if the metric contains a factor $1/(1+r^2)$, the convergence domain is a disk of radius $\sqrt{2}$ centered at the horizon $r_h = 1$, bounded by singularities at $r = \pm i$, thereby restricting reconstruction to within this disk.

A key feature of the aforementioned reconstruction process is that the first $n-2$ equations in $\text{Det}(\vec{\mathcal{M}}^{(n)}(\textbf{g})) = 0$ do not contribute to the metric reconstruction. As detailed in Section \ref{sec_mu_polynomial_constraint}, these $n-2$ redundant equations instead yield $n-2$ polynomial constraints solely in terms of $\mu$. 



\section{Extending the reconstruction method to broader scenarios}\label{sec_reconstruct_extend}

In Section \ref{sec_general_reconstruct}, we introduced our reconstruction method and applied it to planar symmetric static black holes in arbitrary spatial dimension $d$, coupled to a probe massless scalar field. In this section, we extend the applicability of the method in several directions. First, in subsection \ref{subsec_massive_reconstruction} we generalize it to cases with a massive probe scalar field. Then, in subsection \ref{subsec_tensor_channel}, we show that the method remains valid beyond the probe limit by applying it to the tensor sector of gravitational perturbations. In subsection \ref{subsec_hyper_violate}, we apply our method to backgrounds with Lifshitz scaling and hyperscaling violation. Finally, in subsection \ref{subsec_TTbar}, we demonstrate that the reconstruction remains applicable when the boundary CFT is modified by a $T\bar{T}$ deformation.

\subsection{Reconstructing bulk metric coupled to a probe massive scalar field}\label{subsec_massive_reconstruction}

In this section, we extend our reconstruction method to planar-symmetric static black holes in arbitrary spatial dimension $d$, described by the metric \eqref{equ_general_metric_d}, now coupled to a probe scalar field $\phi$ of mass $m$ obeying the Klein–Gordon equation $(\nabla^{2} + m^2)\phi = 0$. Substituting the metric \eqref{equ_general_metric_d} into the Klein–Gordon equation yields the equation of motion for $\varphi$, the Fourier mode of $\phi$, in the following form:

\begin{equation}\label{equ_KGequation_General_d}
\begin{aligned}
&\frac{g_{vv}(r)}{g_{vr}(r)^{2}}\varphi^{\prime\prime}(r)+\left (-\frac{2 i\omega}{g_{vr}(r)} + \frac{d g_{vv}(r)}{rg_{vr}(r)^{2}} - \frac{g_{vv}(r) g_{vr}^{\prime}(r)}{g_{vr}(r)^{3}} + \frac{g_{vv}^{\prime}(r)}{g_{vr}(r)^{2}}\right)\varphi^{\prime}(r)\\
&- \left (m^{2}+\frac{\mu}{r^{2}}+\frac{i d\omega}{r g_{vr}(r)}\right)\varphi (r)=0.\\
\end{aligned}
\end{equation}
Similar to its massless counterpart in Eq. \eqref{equ_KGequation_General_d_mass_zero}, the corresponding determinant equation $\text{Det}(\mathcal{M}^{(n)}(\boldsymbol{\mu})) = 0$ for the massive scalar field also yields a degree-$n$ polynomial in $\mu$, as described by Eq. \eqref{equ_mu_order_n_vieta}. Consequently, we can directly follow the reconstruction procedure outlined in Section \ref{sec_general_reconstruct}.

For $n = 1$, the expansion coefficients $g_{vv_{1}}$ and $g_{vr_{0}}$ are obtained by combining $\text{Det}(\vec{\mathcal{M}}^{(1)}(\textbf{g})) = 0$ with the temperature relation in Eq. \eqref{equ_Tb_Th}, resulting in:

\begin{equation}\label{equ_sol_gvv1_gvr0_d_mass}  
g_{vv_1} = \frac{2 d \omega_1^2}{m^2 + E_{1}(\mu)}, \quad g_{vr_0} = -\frac{i d \omega_1}{m^2 + E_{1}(\mu)}.  
\end{equation}

For $n = 2$, solving $\text{Det}(\mathcal{M}^{(2)}_{1}(\mathbf{g})) = 0$ and $\text{Det}(\mathcal{M}^{(2)}_{2}(\mathbf{g})) = 0$ yields the second-order expansion coefficients:
\begin{equation}\label{equ_sol_gvv2_gvr1_d_mass}
\begin{aligned}
&g_{vv_{2}}=\frac{d \omega_1^2}{4 \left (E_{1}(\mu) + m^2\right)^3}(8 d m^2 E_{1}(\mu) - 3 d m^2 E_{2}(\mu) + 8 d E_{1}(\mu)^2 - 4 d E_{1}(\mu) E_{2}(\mu) \\
&+ d E_{2}(\mu^2) + 20 m^2 E_{1}(\mu) + 12 E_{1}(\mu)^2 + d m^4 + 8 m^4),\\
&g_{vr_{1}}=\frac{-i d \omega_1}{4 \left (E_{1}(\mu) + m^2\right)^3}(4 d m^2 E_{1}(\mu) - d m^2 E_{2}(\mu) + 4 d E_{1}(\mu)^2 - 2 d E_{1}(\mu) E_{2}(\mu)\\
&+ d E_{2}(\mu^2) + 4 m^2 E_{1}(\mu) + 4 E_{1}(\mu)^2 + d m^4).\\
\end{aligned}
\end{equation}
The mass of the bulk scalar field $m$ appearing in Eqs. \eqref{equ_sol_gvv1_gvr0_d_mass} and \eqref{equ_sol_gvv2_gvr1_d_mass} can be replaced, via the holographic dictionary, by the scaling dimension $\Delta$ of the dual boundary operator. Remarkably, this $m$–$\Delta$ relation can also be extracted directly from the pole-skipping data, as shown in Appendix \ref{App_a_equal_1}.

For $n > 2$, following the reasoning in Section \ref{subsec_introduce_linear_mu_constraints}, $g_{vv_n}$ and $g_{vr_{n-1}}$ can again be uniquely determined by solving the pair of equations $\text{Det}(\mathcal{M}^{(n)}_{n-1}(\mathbf{g})) = 0$ and $\text{Det}(\mathcal{M}^{(n)}_{n}(\mathbf{g})) = 0$. The resulting expressions depend solely on boundary data: $\omega_1$, $E_m(\mu^{m-1})$, $E_m(\mu^m)$, and the additional parameter $\Delta$. The explicit forms of $g_{vv_n}$ and $g_{vr_{n-1}}$ for $n > 2$ are too lengthy to display here.

\subsection{Reconstructing bulk metric beyond probe limit}\label{subsec_tensor_channel}

Thus far, we have applied our reconstruction method exclusively to backgrounds coupled with a probe scalar field, where the background metric is unaffected by matter distributions. In this section, we aim to generalize our approach beyond the probe limit.

In general, solving the linearized perturbation equations beyond the probe limit is challenging, as various metric and matter field components become coupled. However, in some cases, this complexity can be significantly reduced by dividing the perturbation modes and their corresponding equations into decoupled sectors, each governed by a Klein-Gordon equation with a potential $V(r)$, commonly referred to as master equations, in terms of gauge-invariant master fields $\Phi(r)$ \cite{kodama_2003_Master_fields_0, kodama_2004_master_fields_1, jansen_2019_Master_fields_KG_type}.

In Einstein gravity, for black holes with a maximally symmetric $d$-dimensional spatial part (with $d>2$), the coupled linearized perturbation equations can generally be decoupled into three sectors: tensor, vector, and scalar, based on their transformation properties under the little group $\text{SO}(d-1)$ \cite{Kovtun_2005_QNM_holography}. Among these, the tensor sector is particularly simple, involving only tensor perturbation modes. The corresponding equations fully decouple and reduce to a single master equation---namely, the massless Klein-Gordon equation \eqref{equ_KGequation_General_d_mass_zero} introduced in Section \ref{sec_general_reconstruct}. We can exploit this simplicity in the tensor sector to implement our reconstruction method free from interactions between different perturbation modes. Consequently, our reconstruction method remains applicable beyond the probe limit.

\subsection{Reconstructing bulk metric with Lifshitz scaling and hyperscaling violation}\label{subsec_hyper_violate}

In this subsection, we focus on reconstructing bulk geometries exhibiting Lifshitz scaling and hyperscaling violation \cite{kachru_2008_gravity_dual_Lifshitz_field_theory, dong_2012_Hyperscaling_violation_original_paper}, characterized by the following metric:

\begin{equation}\label{equ_hyperscaling_metric}
ds^2 = -g_{vv}(r) dv^2 +2 g_{vr}(r) dvdr + g_{xx}(r) d\vec{x}^2,
\end{equation} 
which is covariant under the following scaling transformation:

\begin{equation}\label{equ_hyper_violation_scaling}
\begin{aligned} 
g_{vv} &\to \lambda^{2\mathbf{z} - \frac{2\theta}{d}} g_{vv}, \quad g_{vr} \to \lambda^{\mathbf{z} - \frac{2\theta}{d} - 1} g_{vr}, \quad g_{xx} \to \lambda^{2 - \frac{2\theta}{d}} g_{xx},\\
x_i &\to \lambda^{-1} x_i, \quad v \to \lambda^{-\mathbf{z}} v, \quad r \to \lambda r, \quad ds \to \lambda^{-2\theta/d} ds. \\ 
\end{aligned}
\end{equation}
In this transformation, $\mathbf{z}$ is the dynamical exponent, $\theta$ is the hyperscaling violation exponent, and $d$ represents the spatial dimension. Without loss of generality, we fix the metric element $g_{xx}(r)$ as $g_{xx}(r) = r^{2 - \frac{2\theta}{d}}$, leaving $g_{vr}(r)$ and $g_{vv}(r)$ unspecified in Eq. \eqref{equ_hyperscaling_metric} and assuming a near-horizon expansion as in Eq. \eqref{equ_gvvgvr_expansion}.


Similar to Section \ref{sec_general_reconstruct}, we consider a massless probe scalar field $\phi$ (and its Fourier mode $\varphi$) that satisfies the Klein-Gordon equation analogous to Eq. \eqref{equ_KGequation_General_d}:

\begin{equation}\label{equ_KGequation_hyperviolation_d}
\begin{aligned}
&\frac{g_{vv}(r)}{g_{vr}(r)^{2}}\varphi^{\prime\prime}(r)+\left (-\frac{2 i\omega}{g_{vr}(r)} + \frac{(d-\theta) g_{vv}(r)}{rg_{vr}(r)^{2}} - \frac{g_{vv}(r) g_{vr}^{\prime}(r)}{g_{vr}(r)^{3}} + \frac{g_{vv}^{\prime}(r)}{g_{vr}(r)^{2}}\right)\varphi^{\prime}(r)\\
&- \left (\frac{\mu}{r^{2-\frac{2\theta}{d}}}+\frac{i (d-\theta)\omega}{r g_{vr}(r)}\right)\varphi (r)=0.\\
\end{aligned}
\end{equation}
This equation is invariant under the scaling transformation:

\begin{equation}\label{equ_hyper_violation_scaling_KGequation}
\mu \to \lambda^2 \mu, \quad \omega \to \lambda^{\mathbf{z}} \omega, \quad r \to \lambda r, \quad g_{vv} \to \lambda^{2\mathbf{z} - \frac{2\theta}{d}} g_{vv}, \quad g_{vr} \to \lambda^{\mathbf{z} - \frac{2\theta}{d} - 1} g_{vr}, 
\end{equation}
allowing us to rescale the horizon location to unity. As before, the corresponding determinant equation $\text{Det}(\mathcal{M}^{(n)}(\boldsymbol{\mu})) = 0$ yields a degree-$n$ polynomial in $\mu$.

Then, similar to the procedure in Section \ref{sec_general_reconstruct}, we can solve for $g_{vv_{1}}$ and $g_{vr_{0}}$ as:

\begin{equation}\label{equ_sol_hyper_violatoin_gvv1_gvr0_d}
g_{vv_1}=\frac{2 (d-\theta) \omega_1^2}{E_{1}(\mu)}, \quad g_{vr_0}=-\frac{i (d-\theta) \omega_1}{E_{1}(\mu)}.
\end{equation}
For $n = 2$, solving $\text{Det}(\mathcal{M}^{(2)}_{1}(\mathbf{g})) = 0$ and $\text{Det}(\mathcal{M}^{(2)}_{2}(\mathbf{g})) = 0$ gives out $g_{vv_2}$ and $g_{vr_{1}}$:

\begin{equation}\label{equ_sol_hyper_violatoin_gvv2_gvr1_d}
    \begin{aligned}
        &g_{vv_{2}}=\frac{\omega_1^2 (d - \theta)}{4 d E_{1}(\mu)^3}\left(\left(4 d (2 d + 3) - 8 (d + 2) \theta \right)E_{1}(\mu)^2 + 4 d (\theta - d)E_{2}(\mu) E_{1}(\mu) + d (d - \theta)E_{2}(\mu^2)\right),\\
        &g_{vr_{1}}=- \frac{i (d - \theta) \omega_1}{4 d E_{1}(\mu)^3}\left(d (1 + d) - (2 + d) \theta \right) E_{1}(\mu)^2 + 2 d (\theta - d) E_{2}(\mu) E_{1}(\mu) + d (d - \theta) E_{2}(\mu^2).\\
    \end{aligned}
\end{equation}

Note that the explicit $\theta$ dependence in Eqs. \eqref{equ_sol_hyper_violatoin_gvv1_gvr0_d} and \eqref{equ_sol_hyper_violatoin_gvv2_gvr1_d} arises from the ansatz $g_{xx}(r) = r^{2 - \frac{2\theta}{d}}$. If we instead fix $g_{xx}(r) = r^{2}$, the expressions for $g_{vv_1}$, $g_{vr_0}$, $g_{vv_2}$, and $g_{vr_1}$ simplify to those in Eqs. \eqref{equ_sol_gvv1_gvr0_d} and \eqref{equ_sol_gvv2_gvr1_d}, where all $\theta$ dependence is implicitly encoded within the $E_{n}(\mu^{m})$ terms.

For $n > 2$, the derivation in Appendix \ref{App_linearity_prove} also applies to Eq. \eqref{equ_KGequation_hyperviolation_d}. Consequently, the coefficients $g_{vv_n}$ and $g_{vr_{n-1}}$ can be determined by solving $\text{Det}(\mathcal{M}^{(n)}_{n-1}(\mathbf{g})) = 0$ and $\text{Det}(\mathcal{M}^{(n)}_{n}(\mathbf{g})) = 0$, following the same procedure outlined in Section \ref{subsec_introduce_linear_mu_constraints}, which we do not repeat here.

\subsection{Reconstructing bulk metric with $T\bar{T}$-deformed boundary CFT}\label{subsec_TTbar}

In this subsection, we demonstrate that our reconstruction method remains valid when applied to $T\bar{T}$-deformed boundary CFT. Specifically, we focus on the $2d$ CFT that is dual to the BTZ black holes, as discussed in Section \ref{sec_PS_review}, now subjected to the $T\bar{T}$ deformation.\footnote{The two-point and higher-point correlation functions for the scalar operator and the energy-momentum tensor in the BTZ black hole with a finite cutoff have been calculated in \cite{he_2024_TTbar_BTZ_scalar, he_2024_TTbar_BTZ_stresstensor}.}

The $T\bar{T}$ deformation, first introduced in \cite{cavaglia_2016_TbarT-deformation_original_2a, smirnov_2017_TbarT-Deformation_original_1}, is a notable example of an irrelevant and solvable deformation of $2d$ CFTs. This deformation is driven by the composite operator $\mathcal{O}_{T\bar{T}}$ in terms of stress-energy tensor $T_{\alpha\beta}$ as follows:

\begin{equation}\label{equ_TTbar_operator}
\mathcal{O}_{T\bar{T}}=\frac{1}{2}\left (T^{\alpha\beta}T_{\alpha\beta}-T_{\alpha}^{\alpha}T_{\beta}^{\beta}\right).
\end{equation}
Generally, the deformed action $S_{\lambda}$ can be written as:

\begin{equation}\label{equ_TTbar_action}
\frac{dS_\lambda}{d\lambda}=\int d^{2}x\sqrt{g}\mathcal{O}_{T\bar{T}},
\end{equation}
where $\lambda$ is the coupling of the deformation and $S_{\lambda=0}$ represents the action of the original $2d$ CFT.

The $T\bar{T}$ deformation provides a rare example of an exactly solvable irrelevant deformation that can be understood from both a field-theoretic and holographic perspective. In the context of holography, the $T\bar{T}$ deformation is understood as introducing a geometric cutoff in the dual gravity description \cite{mcgough_2018_TbarT-deformation_Holographya}. This cutoff removes the asymptotic region of the $AdS_3$ space and places the deformed QFT on a Dirichlet wall at a finite radial distance $r = r_c$ in the bulk. Consider pure $\text{AdS}_3$ gravity with all matter sources turned off, the relationship between the coupling parameter $\lambda$ in the deformed action \eqref{equ_TTbar_action} and the cutoff radius $r_c$ is given by

\begin{equation}\label{equ_TTbar_coupling}
\lambda = \frac{2\pi G}{r_c^2} = \frac{3\pi}{c} \frac{1}{r_c^2},
\end{equation}
with $G$ being the Newton's constant and $c$ the central charge of the boundary CFT.

With this background established, we consider the BTZ black hole as the bulk geometry, coupled solely to a probe scalar field $\phi(r)$ of mass $m$, as discussed in Section \ref{sec_PS_review}. By imposing a Dirichlet boundary condition at $r = r_c$ on both the original BTZ black hole geometry and the probe scalar field, the dual effective field theory on the cutoff surface can be defined through the corresponding flow equation (see Eq. 2.6 in \cite{hartman_2019_TbarT-Deformation_correlation_function}). Utilizing the holographic dictionary at finite cutoff outlined in \cite{hartman_2019_TbarT-Deformation_correlation_function}, we analytically derive the scalar retarded Green’s function $\mathcal{G}^{\mathcal{O}}_{z_c}$ with $z_c = \frac{1}{r_c}$. Due to its length, the explicit expression is presented in Appendix \ref{App_TTbar_GR} as Eq. \eqref{equ_TTbar_GR}.

\begin{figure}[h!] \centering \includegraphics[width=1\textwidth]{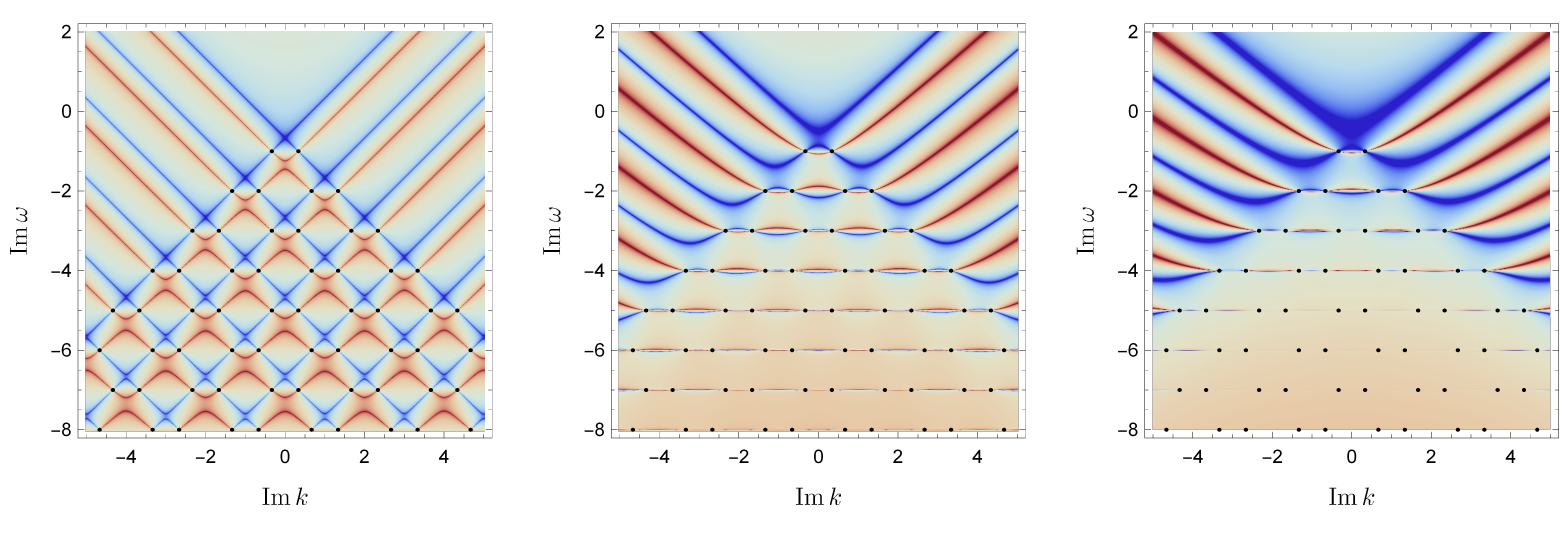} \caption{\label{fig_TTbar_GR} The figure displays heatmaps showing the values of $\log|\mathcal{G}^{\mathcal{O}}_{z_c}(\mathrm{Im}\omega,\mathrm{Im}k,z_{c})|$ for different values of $z_{c}$: $\frac{5}{1000}$, $\frac{3}{10}$, and $\frac{5}{10}$ (with $T=\frac{1}{2\pi}$ and $\Delta=\frac{4}{3}$). The red and blue lines represent the poles and zeroes of Eq. \eqref{equ_TTbar_GR}, respectively. The black dots indicate the pole-skipping points obtained from the near-horizon analysis of the BTZ black hole described by Eq. \eqref{equ_intersections_poles_zeroes_BTZ}. In all panels, the black dots align precisely with the intersections of poles and zeroes of $\mathcal{G}^{\mathcal{O}}_{z_c}(\omega,k,z_{c})$, implying that the pole-skipping points remain invariant under $T\bar{T}$ deformation.} 
\end{figure}

\begin{figure}[h!] \centering \includegraphics[width=1\textwidth]{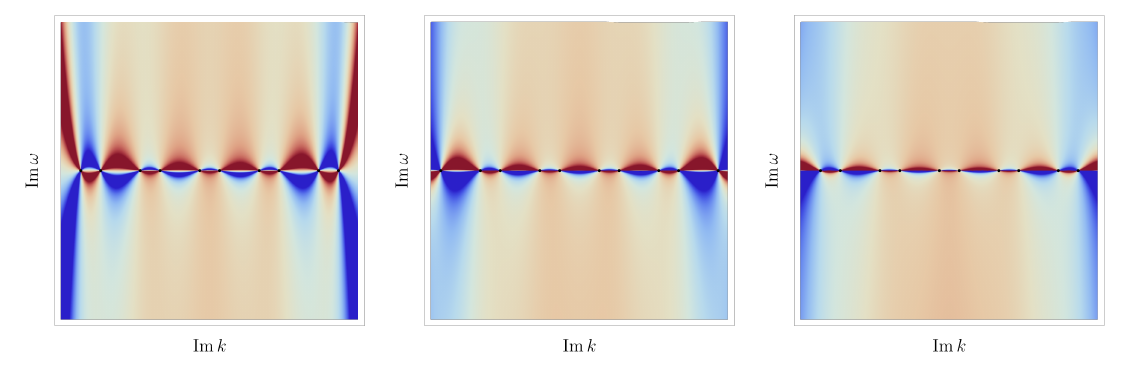} \caption{\label{fig_TTbar_GR_enlarge} From left to right, the higher-resolution version of the third panel of Figure \ref{fig_TTbar_GR} around different values of $\omega$: $-5i$, $-6i$, and $-7i$.} \end{figure}

The pole-skipping structure of $\mathcal{G}^{\mathcal{O}}_{z_c}(\omega, k, z_c)$ is illustrated in Figure \ref{fig_TTbar_GR} and Figure \ref{fig_TTbar_GR_enlarge}. Remarkably, although the poles and zeroes shift significantly with varying radial cutoffs $z_c$, the locations of all pole-skipping points in $\mathcal{G}^{\mathcal{O}}_{z_c}(\omega, k, z_c)$ remain identical to those of the BTZ scalar Green’s function $\mathcal{G}^{\mathcal{O}}_{R}(\omega, k)$ \eqref{equ_BTZ_GR}, explicitly given by Eq. \eqref{equ_intersections_poles_zeroes_BTZ}.

If our reconstruction method is valid in the context of $T\bar{T}$ deformations, then the identical pole-skipping data from $\mathcal{G}^{\mathcal{O}}_{z_c}(\omega, k, z_c)$ with different $z_c$ should yield the same $g_{vv_n}$ and $g_{vr_{n-1}}$ as in the original BTZ case. This is indeed consistent with the fact that the $T\bar{T}$ deformation is irrelevant, thereby leaving the near-horizon expansion coefficients  $g_{vv_{n}}$ and $g_{vr_{n-1}}$ unaffected. Consequently, our reconstruction method is indeed applicable to theories deformed by $T\bar{T}$ deformation.

To conclude this section, we summarize our reconstruction method together with its assumptions and scope of applicability. The method reconstructs $g_{vv_{n}}$ and $g_{vr_{n-1}}$ within the general metric ansatz \eqref{equ_general_metric_d}. The key assumption is of course the existence of a boundary QFT exhibiting the pole-skipping phenomenon. In addition, we assume that the bulk equations of motion take the form $(\nabla^{2}+m^{2})\phi(r)=0$, with $m$ possibly vanishing. This Klein–Gordon–type equation dictates the structure of the pole-skipping points and allows the reconstruction to be completed by solving the associated linear equations $\text{Det}(\mathcal{M}^{(n)}(\mathbf{g}))=0$. Finally, the maximal bulk spacetime recoverable by our method is bounded by the nearest singularities in the complex-$r$ plane relative to $r = r_{h}$.

\section{Reinterpret Einstein equation via pole-skipping points}\label{sec_Einstein_equation_PS}

In previous sections, we introduced how bulk black hole geometries can be reconstructed from boundary pole-skipping data. This reconstruction naturally extends to any geometric quantity that depends on the metric, allowing such quantities to be reinterpreted in terms of pole-skipping data.

In this section, we focus on one such quantity: the vacuum Einstein equation with a negative cosmological constant, given by
\begin{equation}\label{equ_vacuum_Einstein_equation}
E_{\mu\nu} \equiv R_{\mu\nu}-\frac{1}{2}g_{\mu\nu}R+\Lambda g_{\mu\nu}=0,
\end{equation}
where the cosmological constant is $\Lambda = -\frac{d(d+1)}{2}$, with the AdS radius set to unity. To align with the reconstruction framework of Section \ref{sec_general_reconstruct}, we consider a static, planar-symmetric black hole ansatz described by the metric \eqref{equ_general_metric_d} as a solution to Eq. \eqref{equ_vacuum_Einstein_equation}.

We then expand $E_{\mu\nu}$ near the horizon as
\begin{equation}\label{equ_Emunu_near-horizon_expand}
E_{\mu\nu} = E_{\mu\nu_0} + E_{\mu\nu_1}(r - 1) + E_{\mu\nu_2}(r - 1)^2 + \ldots
\end{equation}
Since $E_{\mu\nu}$ contains at most second-order derivatives of the metric, each $E_{\mu\nu_n}$ depends at most on $g_{vv_{n+2}}$ and $g_{vr_{n+1}}$.

At zeroth order, three nontrivial components appear: $E_{vr_0} = 0$, $E_{rr_0} = 0$, and $E_{xx_0} = 0$, whose explicit forms are:
\begin{equation}\label{equ_Evr0_Err0_Exx0_in_gvv_gvr}
\begin{aligned}
&E_{vr_{0}}=\frac{d \left ( (1+d) g_{vr_0}^2 - g_{vv_1} \right)}{2 g_{vr_0}}=0,\\
&E_{rr_{0}}=\frac{d g_{vr_1}}{g_{vr_0}}=0,\\
&E_{xx_{0}}=-\frac{1}{2} d (1 + d) - \frac{g_{vr_1} g_{vv_1}}{2 g_{vr_0}^3} + \frac{(d-1) g_{vv_1} + g_{vv_2}}{g_{vr_0}^2}=0.\\
\end{aligned}
\end{equation}
Substituting the reconstructed metric components from Eqs. \eqref{equ_sol_gvv1_gvr0_d} and \eqref{equ_sol_gvv2_gvr1_d}, these equations can be fully recast in terms of the pole-skipping data:

\begin{equation}\label{equ_Evr0_Err0_Exx0_PS}
\begin{aligned}
&E_{vr_{0}}=\frac{d \omega_1}{E_{1}(\mu)} \left (2 E_{1}(\mu) + d^2 + d\right)=0,\\
&E_{rr_{0}}=d \left (4 (d+1) E_{1}(\mu) - 2 d E_{2}(\mu) + \frac{d E_{2}(\mu^2)}{E_{1}(\mu)}\right)=0,\\
&E_{xx_{0}}=6 E_{1}(\mu)-E_{2}(\mu) + d^2 + d =0.\\
\end{aligned}
\end{equation}

This recasts the near-horizon expansion of the vacuum Einstein equation as a set of constraint equations on the pole-skipping data. Solving them yields:
\begin{equation}\label{equ_Evr0_Err0_Exx0_solve_Enmum}
E_{1}(\mu) = -\frac{1}{2} (d +d^2), \quad E_{2}(\mu) = -2 (d + d^2), \quad E_{2}(\mu^2) = (d-1)\, d\, (1 + d)^{2}.
\end{equation}

Proceeding to first order, after substituting all known solutions for $g_{vv_n}$ and $g_{vr_{n-1}}$ with $n < 4$, along with Eq. \eqref{equ_Evr0_Err0_Exx0_solve_Enmum}, only two nontrivial components remain: $E_{rr_1} = 0$ and $E_{xx_1} = 0$. These translate into two constraint equations for $E_3(\mu^2)$ and $E_3(\mu^3)$:
\begin{equation}\label{equ_Err1_Exx1_solve_PS}
\begin{aligned}
&E_{rr_{1}} =3 d E_{3}(\mu^2) + \frac{2 E_{3}(\mu^3)}{1 + d} - 6 d (1 + d)^3 (3 d - 2)=0,\\
&E_{xx_{1}}=-\frac{1}{8} d (1 + d)(39 d - 20) + \frac{E_{3}(\mu^2)}{2 (1 + d)}=0.\\
\end{aligned}
\end{equation}
Solving these equations gives:
\begin{equation}\label{equ_Err1_Exx1_solve_Enmum}
E_{3}(\mu^{2})=\frac{1}{4} d (d+1)^2 (39 d-20), \quad E_{3}(\mu^{3})=-\frac{3}{8} d (d+1)^3 \left(15 d^2-28 d+16\right).
\end{equation}
This reinterpretation of the near-horizon Einstein equations in terms of pole-skipping data naturally extends to higher orders, though we do not present those results explicitly. In general, for each $n > 0$, the $n^{\text{th}}$-order components of the Einstein equation yield two algebraic equations that uniquely determine $E_{n+2}(\mu^{n+2})$ and $E_{n+2}(\mu^{n+1})$. This is expected, as the vacuum Einstein equation \eqref{equ_vacuum_Einstein_equation} completely determines the background black hole geometry. Consequently, all pole-skipping points associated with the Klein-Gordon equation \eqref{equ_KGequation_General_d_mass_zero} must be encoded in the near-horizon structure.

However, the expressions for $E_n(\mu^m)$ with $m = 1, \ldots, n-2$ are not directly obtained through this recursive procedure, even though all pole-skipping data are, in principle, determined. In the next section, we show that these $n-2$ elementary symmetric $\mu$-polynomials are in fact uniquely fixed by a set of $n-2$ universal homogeneous polynomial identities in $\mu$.

To conclude this section, we note that when matter is present and described by an energy-momentum tensor $T_{\mu\nu}$, the above derivation provides a reinterpretation of its near-horizon expansion in terms of boundary pole-skipping data.

\section{$\mu$-polynomial constraints derived from master equations}\label{sec_mu_polynomial_constraint}

Only the last two equations in $\text{Det}(\vec{\mathcal{M}}^{(n)}(\textbf{g})) = 0$ are used to reconstruct the bulk geometry, as discussed in Sections \ref{sec_general_reconstruct} and \ref{sec_reconstruct_extend}, and to reinterpret the near-horizon Einstein equations in Section \ref{sec_Einstein_equation_PS}. The remaining $n-2$ equations may seem irrelevant. However, in this section, we demonstrate that they give rise to $n-2$ independent homogeneous polynomial constraints on $\mu$, of degrees $1, 2, \ldots, n-2$, respectively.

To explore the widest scope in which these polynomial constraints remain valid, we consider the master equation in a generalized Klein-Gordon form:
\begin{equation}
\label{equ_masterequation_V}
(\nabla^2 + V (r))\Phi (r) = 0.
\end{equation}
where $\Phi$ (or its Fourier mode $\Psi$) denotes a gauge-invariant master field \cite{kodama_2003_Master_fields_0, kodama_2004_master_fields_1, jansen_2019_Master_fields_KG_type}, as introduced in Section \ref{subsec_tensor_channel}. The potential $V(r)$ depends on the specifics of the theory, and the covariant derivative $\nabla$ is defined with respect to the background metric \eqref{equ_general_metric_d}. The explicit form of the master equation \eqref{equ_masterequation_V} in this background reads:

\begin{equation}\label{equ_master_equation_General}
\begin{aligned}
&\frac{g_{vv}(r)}{g_{vr}(r)^{2}}\Psi^{\prime\prime}(r) + \left(-\frac{2 i\omega}{g_{vr}(r)} + \frac{d g_{vv}(r)}{r g_{vr}(r)^{2}} - \frac{g_{vv}(r) g_{vr}^{\prime}(r)}{g_{vr}(r)^{3}} + \frac{g_{vv}^{\prime}(r)}{g_{vr}(r)^{2}}\right)\Psi^{\prime}(r) \\
&- \left(V(r) + \frac{\mu}{r^{2}} + \frac{i d \omega}{r g_{vr}(r)}\right)\Psi(r) = 0,
\end{aligned}
\end{equation} 
We assume that the potential $V(r)$ admits a Taylor expansion near the horizon, analogous to those of $g_{vv}(r)$ and $g_{vr}(r)$ in Eq. \eqref{equ_gvvgvr_expansion}:
\begin{equation}\label{equ_V_expansion}
V(r) = V_0 + V_1 (r - 1) + V_2 (r - 1)^2 + \cdots
\end{equation}
We can then expand the master equation \eqref{equ_master_equation_General} at the horizon up to the $n^{\text{th}}$ order, giving rise to the determinant equation $\text{Det}(\mathcal{M}^{(n)}(\boldsymbol{\mu})) =0$. Here, we assume that
\begin{Ass}\label{ass_1}
$\text{Det}(\mathcal{M}^{(n)}(\boldsymbol{\mu}))=0$ derived from master equation \eqref{equ_master_equation_General} is a degree-$n$ polynomial in $\mu$, taking the form given in Eq. \eqref{equ_mu_order_n_vieta}.
\end{Ass}
Equivalently, this assumption asserts that $\text{Det}(\mathcal{M}^{(n)}(\boldsymbol{\mu})) = 0$ admits exactly $n$ roots in $\mu$. Combined with the linearity proof provided in Appendix \ref{App_linearity_prove}, this ensures that the key result established earlier for the Klein-Gordon equation \eqref{equ_KGequation_General_d_mass_zero} remains valid in the current setting: namely, the pair of equations $\text{Det}(\mathcal{M}^{(n)}_{n-1}(\mathbf{g})) = 0$ and $\text{Det}(\mathcal{M}^{(n)}_{n}(\mathbf{g})) = 0$ are sufficient to uniquely determine $g_{vv_{n}}$ and $g_{vr_{n-1}}$, while the remaining $n-2$ equations in $\text{Det}(\mathcal{M}^{(n)}(\boldsymbol{\mu})) = 0$ do not contribute to the reconstruction.

Moreover, Assumption \ref{ass_1} constrains the potential $V(r)$ such that, in its near-horizon expansion, the coefficient $V_m$ contains powers of $\mu$ no higher than $m+1$ for any $m\geq0$. The full implication of this assumption for the general form of $V(r)$, however, remains unclear. A plausible candidate under this assumption is any $V(r)$ that is at most linear in $\mu$.

Note that the following analysis does not adhere to the framework of our previously discussed reconstruction method. In particular, we do not assume the existence of a boundary QFT dual to the bulk spacetime under consideration. In such scenarios, the pole-skipping points $(\omega_n, \mu_{n,q})$ represent special points where two ingoing solutions coexist at the horizon, disregarding their field theory interpretations. Indeed, the pole-skipping phenomenon can be present in classical gravity without a holographic dual boundary QFT. For example, pole-skipping points of four-dimensional massive black holes have been investigated purely from the perspective of classical gravity in \cite{grozdanov_2023_Pole_skipping_hyperbolic_sphereical_flat}.

\subsection{$\mu$-polynomial constraints at $n=3$}\label{subsec_linear_mu_constraints_general}

We begin with examining $\text{Det}(\mathcal{M}^{(3)}_{1}(\textbf{g})) = 0$, or equivalently $E_{3}(\mu) - \frac{v_{3,2}}{v_{3,3}} = 0$, derived from the master equation \eqref{equ_master_equation_General}, which takes the explicit form: 

\begin{equation}\label{equ_linear_mu_constrain_origin}
E_{3}(\mu) + \frac{4 g_{vr_1} g_{vv_1}}{g_{vr_0}^3} + \frac{\left (8 + \frac{3 d}{2}\right) g_{vv_1} - 8 g_{vv_2}}{g_{vr_0}^2} + 3 V_0=0.
\end{equation}
This equation does not involve $g_{vv_3}$ or $g_{vr_2}$ and therefore does not contribute to their reconstruction, as previously noted. Following the recursive procedure in Section \ref{sec_general_reconstruct}, the solutions for $g_{vv_1}$, $g_{vr_0}$ and $g_{vv_2}$, $g_{vr_1}$ are readily obtained as:

\begin{equation}\label{equ_sol_gvv1_gvr0_V}
    g_{vv_1} = \frac{2 d \omega_1^2}{E_{1}(\mu) + V_0}, \quad g_{vr_0} = -\frac{i d \omega_1}{E_{1}(\mu) + V_0},
\end{equation}
and
\begin{equation}\label{equ_sol_gvv2_gvr1_V}
\begin{aligned}
g_{vv_2} &= \frac{d \omega_1^2}{4 \left(E_{1}(\mu) + V_0\right)^3} \Big( 4 (3 + 2 d) E_{1}(\mu)^2 - 4 d E_{1}(\mu) E_{2}(\mu) + d E_{2}(\mu^2) \\
&\quad + V_0 \left(4 (5 + 2 d) E_{1}(\mu) - 3 d E_{2}(\mu) + (8 + d) V_0 \right) - 2 \left(E_{1}(\mu) + V_0\right) V_1 \Big), \\[1ex]
g_{vr_1} &= -\frac{i d \omega_1}{4 \left(E_{1}(\mu) + V_0\right)^3} \Big( 4 (1 + d) E_{1}(\mu)^2 - 2 d E_{1}(\mu) E_{2}(\mu) + d E_{2}(\mu^2) \\
&\quad + V_0 \left(4 (1 + d) E_{1}(\mu) - d E_{2}(\mu) + d V_0 \right) - 2 \left(E_{1}(\mu) + V_0\right) V_1 \Big),
\end{aligned}
\end{equation}
where the frequency $\omega_1$ should be understood as $\omega_1 = -i 2\pi T_h$. It is important to emphasize that these expressions should not be viewed as a boundary-to-bulk reconstruction, since the potential coefficients $V_i$ generally lack a clear boundary interpretation. Rather, they represent a reformulation of the near-horizon metric components in terms of other bulk quantities. 

Substituting these solutions back into Eq. \eqref{equ_linear_mu_constrain_origin} simplifies it to a homogeneous polynomial identity purely in terms of $\mu$:
\begin{equation}\label{equ_linear_mu_constrain_intermediate}
E_{3}(\mu)-4E_{2}(\mu)+5E_{1}(\mu)=0,
\end{equation} 
or more intuitively,
\begin{equation}\label{equ_linear_mu_constrain_3}
\mu_{3,1} + \mu_{3,2} + \mu_{3,3} - 4 (\mu_{2,1}+\mu_{2,2})+5\mu_{1,1}=0.
\end{equation}
Remarkably, all dependence on $d$ and $V_i$ in Eqs. \eqref{equ_sol_gvv1_gvr0_V} and \eqref{equ_sol_gvv2_gvr1_V} cancels out upon substitution into Eq. \eqref{equ_linear_mu_constrain_origin}, implying that this identity holds regardless of the specific form of $V(r)$ or the spacetime dimension $d$.  

Its validity can be readily verified in the BTZ case by substituting Eq. \eqref{equ_intersections_poles_zeroes_BTZ} into Eq. \eqref{equ_linear_mu_constrain_3}. This polynomial constraint implies that at $n = 3$, once $\mu_{1,1}$, $\mu_{2,1}$, and $\mu_{2,2}$ are known, only two of the $\mu_{3,q}$ remain independent, matching the number of variables $g_{vv_2}$ and $g_{vr_1}$, while the third constrained by Eq. \eqref{equ_linear_mu_constrain_3}. We therefore refer to identities like Eq. \eqref{equ_linear_mu_constrain_3} as $\mu$-polynomial constraints.

\subsection{$\mu$-polynomial constraints at $n=4$ and $n=5$}\label{subsec_n4n5_mu_constraints_general}

At order $n = 4$, two redundant equations: $\text{Det}(\mathcal{M}^{(4)}_{1}(\mathbf{g})) = 0$ and $\text{Det}(\mathcal{M}^{(4)}_{2}(\mathbf{g})) = 0$ do not contribute to the reformulation of the metric. These equations take the form:

\begin{equation} \label{equ_P4_mu_linear_original}
E_{4}(\mu) + \frac{10 g_{vr_1} g_{vv_1}}{g_{vr_0}^3} + \frac{2 (10 + d) g_{vv_1} - 20 g_{vv_2}}{g_{vr_0}^2} + 4 V_0 = 0,  
\end{equation}

\begin{equation}\label{equ_P4_mu_quadratic_original}
\begin{aligned}
&E_{4}(\mu^2)- \frac{45 g_{vr_1}^2 g_{vv_1}^2}{g_{vr_0}^6} + \frac{6 g_{vv_1} \left(2 g_{vr_2} g_{vv_1} + g_{vr_1} \left(- (22 + 5d) g_{vv_1} + 24 g_{vv_2} \right)\right)}{g_{vr_0}^5} \\
&+ \frac{(-108 - 30d + d^2) g_{vv_1}^2 - 108 g_{vv_2}^2 + 4 g_{vv_1} \left(2 (27 + 5d) g_{vv_2} - 9 g_{vv_3} \right)}{g_{vr_0}^4}\\
&+ \frac{60 g_{vv_2} V_0 - 2 g_{vv_1} \left((20 + 3d) V_0 - 5 V_1 \right)}{g_{vr_0}^2}- \frac{30 g_{vr_1} g_{vv_1} V_0}{g_{vr_0}^3} - 6 V_0^2 = 0,\\
\end{aligned}
\end{equation}
respectively. Again, upon substituting the previously obtained values of $g_{vv_n}$ and $g_{vr_{n-1}}$ for $n = 1, 2, 3$, both equations reduce to homogeneous polynomial identities involving only $\mu$:
\begin{equation}\label{equ_P4_mu_linear}
E_{4}(\mu)-10E_{2}(\mu)+16E_{1}(\mu)=0,
\end{equation}
\begin{equation}\label{equ_P4_mu_quadratic_invariant}
E_{4}(\mu^2)-6 E_{3}(\mu^2)+14 E_{2}(\mu^2)-9 E_{2}(\mu)^2+40 E_{2}(\mu) E_{1}(\mu)-46 E_{1}(\mu)^2=0.
\end{equation}

For convenience, we denote these constraints as $P_n(\mu^m) = 0$, where $n$ is the expansion order and $m$ the degree of the homogeneous polynomial. For instance, the linear constraint \eqref{equ_linear_mu_constrain_3} is denoted as $P_3(\mu) = 0$.

Both $P_4(\mu) = 0$ \eqref{equ_P4_mu_linear} and $P_4(\mu^2) = 0$ \eqref{equ_P4_mu_quadratic_invariant} can be explicitly verified by inserting the pole-skipping locations of the BTZ black hole, as given in Eq. \eqref{equ_intersections_poles_zeroes_BTZ}. As in the $n=3$ case, these two constraints reduce the number of independent $\mu_{4,q}$ to two, matching the number of near-horizon coefficients: $g_{vv_4}$ and $g_{vr_3}$.

At $n=5$, nothing essential changes. After substituting the previously solved expressions for $g_{vv_n}$ and $g_{vr_{n-1}}$ with $n < 5$, the three redundant equations in $\text{Det}(\vec{\mathcal{M}}^{(5)}(\mathbf{g})) = 0$, namely $\text{Det}(\mathcal{M}^{(5)}_{1}(\mathbf{g})) = 0$, $\text{Det}(\mathcal{M}^{(5)}_{2}(\mathbf{g})) = 0$ and $\text{Det}(\mathcal{M}^{(5)}_{3}(\mathbf{g})) = 0$ reduce to three $\mu$-polynomial constraints: $P_{5}(\mu)=0$, $P_{5}(\mu^{2})=0$ and $P_{5}(\mu^{3})=0$, given by

\begin{equation}\label{equ_P5_mu_linear}
E_{5}(\mu) - 20 E_{2}(\mu) + 35 E_{1}(\mu) = 0,
\end{equation}

\begin{equation}\label{equ_quadratic_mu_constrain_5_invariant}
    E_{5}(\mu^2)-315 E_{1}(\mu)^2 + 280 E_{1}(\mu) E_{2}(\mu)- 64 E_{2}(\mu)^2 + 64 E_{2}(\mu^2) - 21 E_{3}(\mu^2)=0,
\end{equation}
and
\begin{equation}\label{equ_P5_cubic_invariant}
\begin{aligned}
&E_{5}(\mu^3)+657 E_{1}(\mu)^3 - 580 E_{1}(\mu)^2 E_{2}(\mu) + 128 E_{1}(\mu) E_{2}(\mu)^2 - 448 E_{1}(\mu) E_{2}(\mu^2) \\
&+ 208 E_{2}(\mu) E_{2}(\mu^2) + 102 E_{1}(\mu) E_{3}(\mu^2) - 48 E_{2}(\mu) E_{3}(\mu^2) + 27 E_{3}(\mu^3) - 8 E_{4}(\mu^3)=0.\\
\end{aligned}
\end{equation}

As in the cases of $n = 3$ and $n = 4$, the polynomial constraints $P_{5}(\mu^{3}) = 0$ \eqref{equ_P5_cubic_invariant}, $P_{5}(\mu^{2}) = 0$ \eqref{equ_quadratic_mu_constrain_5_invariant}, and $P_{5}(\mu) = 0$ \eqref{equ_P5_mu_linear} collectively reduce the number of independent $\mu_{5,q}$ from five to two, precisely matching the number of metric coefficients $g_{vv_{5}}$ and $g_{vr_{4}}$.

Building on the derivations of $P_{n}(\mu^{m})$ for $n = 3$, $4$, and $5$, we now propose a general procedure for constructing all such polynomial constraints $P_{n}(\mu^{m})$ for arbitrary $n$, as detailed in the next subsection.

\subsection{$\mu$-polynomial constraints for arbitrary $n$}\label{subsec_general_n_mu_constraints}

Generally, at expansion order $n$, Assumption \ref{ass_1} guarantees that $\text{Det}(\vec{\mathcal{M}}^{(n)}(\mathbf{g})) = 0$ yields $n$ independent equations. Among these, the final two equations are employed to solve for $g_{vv_{n}}$ and $g_{vr_{n-1}}$, while the first $n - 2$ equations take the explicit form:  

\begin{equation}\label{equ_qth_equation_first_n-2}
\text{Det}(\mathcal{M}^{(n)}_{q}(\mathbf{g})) \equiv E_n (\mu^{q}) - \frac{v_{n,\, n - q}}{v_{n,\, n}} = 0, \quad \text{for } q = 1, 2, \ldots, n - 2.
\end{equation}
The term $\frac{v_{n\, n - q}}{v_{n\, n}}$ is a generally intricate algebraic function depending on $d$, $V_{m-1}$, $g_{vv_m}$, and $g_{vr_{m-1}}$ for $m < n$.

By recursively substituting the known solutions for $g_{vv_m}$ and $g_{vr_{m-1}}$, obtained from the pair of equations $\text{Det}(\mathcal{M}^{(m)}_{m}(\textbf{g})) = 0$ and $\text{Det}(\mathcal{M}^{(m)}_{m-1}(\textbf{g})) = 0$ for all $m < n$, into Eq. \eqref{equ_qth_equation_first_n-2}, each of the $n - 2$ remaining equations simplifies to a homogeneous polynomial identity in $\mu$ alone:
\begin{equation}
    P_n(\mu) = 0,\, P_n(\mu^2) = 0,\, \ldots,\, P_n(\mu^{n - 2}) = 0,
\end{equation}
i.e., a set of $n - 2$ universal polynomial constraints on $\mu$ alone, free of any other quantities. In the context of bulk reconstruction developed in Section \ref{sec_general_reconstruct}, these polynomial constraints remain valid and effectively reduce the number of independent $\mu_{n,q}$ from $n$ to 2, aligning with the number of reconstructed variables $g_{vv_n}$ and $g_{vr_{n-1}}$. While $g_{vv_n}$ and $g_{vr_{n-1}}$ are expressed in terms of $E_n(\mu^{n})$ and $E_n(\mu^{n-1})$, which depend on all $\mu_{n,q}$, the constraints guarantee that specifying any two $\mu_{n,q}$ for each $n > 2$ is sufficient, with the remaining $\mu_{n,q}$ fixed by the $\mu$-polynomial constraints.

With the recursive procedure of obtaining $P_n(\mu^m)$ established, we identify the following structural patterns governing the general form of the $\mu$-polynomial constraints $P_n(\mu^m)$:

\begin{enumerate}
    \item $P_n(\mu^m)$ for $m \leq n - 2$ comprises linear combinations of all possible monomials of the form $\prod_{l=1}^{N} E_{i_l}(\mu^{j_l})$, where $\sum_{l=1}^{N} j_l = m$ and $i_l \leq n$. 
    \item In each $P_n(\mu^m)$, there exists a unique linear monomial $E_n(\mu^m)$ (i.e., with $N = 1$), whose coefficient is fixed to be $1$.
    \item Apart from this distinguished monomial in Rule 2, all other monomials, whether linear or nonlinear, only involve factors $E_{i_l}(\mu^{j_l})$ such that each exponent $j_l$ equals either $i_l$ or $i_l - 1$.
\end{enumerate}

By adhering these rules, the $\mu$-polynomial constraints take the general form: \begin{equation}\label{equ_general_p_nmu^m}
    P_n\left (\mu^m\right)=\sum\limits_{p=1}^{K_m} C_{\left\{\left\{i_{p, 1}, j_{p, 1}\right\},\ldots,\left\{i_{p, N_p}, j_{p, N_p}\right\}\right\}}^{(n, m)} \prod_{q=1}^{N_{p}}E_{i_{p, q}}(\mu^{j_{p, q}}),
\end{equation}
where $C_{\left\{\left\{i_{p,1}, j_{p,1}\right\},\ldots,\left\{i_{p,N_p}, j_{p,N_p}\right\}\right\}}^{(n,m)}$ are undetermined coefficients, and $\sum\limits_{p=1}^{K_m}$ indicates summing over all combinations $\left\{\left\{i_{p,1}, j_{p,1}\right\},\ldots,\left\{i_{p,N_p}, j_{p,N_p}\right\}\right\}$ that satisfy the above rules. For $m$ ranging from 1 to 6, the values of $K_m$ are $3, 6, 11, 21, 37, 66,$ respectively.  

For linear monomials $E_i(\mu^j)$, i.e., when $N_p=1$, the coefficients $C$ can be explicitly determined (by induction) as: 

\begin{equation}\label{equ_single_term_coe}
    C_{\left\{\left\{m+1, m\right\}\right\}}^{(n, m)} = -\binom{n+m}{2m+1}, \quad C_{\left\{\left\{m, m\right\}\right\}}^{(n, m)} = \frac{2m \binom{n+m+1}{2m+3}(2m+3)!}{(n-m)(n+m)(2m+1)!},
\end{equation}
where $\binom{n}{m}$ denotes the binomial coefficient. However, a general formula for general $C_{\{\ldots\}}^{(n,m)}$ remains elusive.

To conclude this section, we conjecture the general formulas for $P_{n}(\mu^{m})$ with $m\leq 4$, based on inductive patterns.

For $P_{n}(\mu)$, we propose the following general formula:
\begin{equation}\label{equ_Pn_mu_linear_explicit}
E_n (\mu) - \binom{n+1}{3} E_2 (\mu)+\frac{40 \binom{n+2}{5}}{(n - 1)(n + 1)} E_1 (\mu) = 0,
\end{equation}
Its validity can be readily verified by setting $n$ to 3, 4, and 5, which unsurprisingly recover $P_{3}(\mu)= 0$ \eqref{equ_linear_mu_constrain_3}, $P_{4}(\mu)= 0$ \eqref{equ_P4_mu_linear}, and $P_{5}(\mu)= 0$ \eqref{equ_P5_mu_linear}, respectively. For arbitrary $n$, the identity \eqref{equ_Pn_mu_linear_explicit} continues to hold when incorporating the BTZ pole-skipping data \eqref{equ_intersections_poles_zeroes_BTZ}, along with all the pole-skipping data presented in Section \ref{sec_example_verify_polynomial_constraint}.\footnote{ 
First, expand the binomial coefficients, cancel the common factor in front of all terms $E_m(\mu^m)$ with $m>1$, and then substitute the specific value of $n$ into them.} A rigorous proof of this general formula is provided in Section \ref{subsec_explain_poly_nearhorizon}.

For $P_{n}(\mu^{2})$, we propose that for any $n$,
\begin{equation}\label{equ_Pn_mu_2_explicit}
    \begin{aligned}
        &E_n (\mu^2)- \binom{n+2}{5} E_3 (\mu^2)+ \frac{168 \binom{n+3}{7}}{(-2 + n)(2 + n)} E_2 (\mu^2) + \frac{20 (2 + n)}{3} \binom{n+1}{5} E_1 (\mu) E_2 (\mu) \\ 
        &-\frac{(2 + n)(7 + 4n)}{3} \binom{n}{4} E_1 (\mu)^2- \frac{1}{3}(7 + 5n) \binom{n+1}{5} E_2 (\mu)^2 = 0. \\
    \end{aligned}
\end{equation}
Setting $n = 4$ and $5$ reproduces the established constraints $P_4(\mu^2) = 0$ \eqref{equ_P4_mu_quadratic_invariant} and $P_5(\mu^2) = 0$ \eqref{equ_quadratic_mu_constrain_5_invariant}. Its validity for other values of $n$ can be confirmed using pole-skipping data from various examples provided in Section \ref{sec_example_verify_polynomial_constraint}.

The general formulas for $P_{n}(\mu^{3})=0$ and $P_{n}(\mu^{4})=0$ are too lengthy to present here and are provided in Appendix \ref{App_Pnmu3_Pnmu4} as Eq. \eqref{equ_Pn_mu_3_explicit} and Eq. \eqref{equ_Pn_mu_4_explicit}. Again, their validity can be verified by incorporating the pole-skipping data from various settings discussed in Section \ref{sec_example_verify_polynomial_constraint}.

\subsection{Proof of the General Formula for $P_n(\mu)=0$
}\label{subsec_explain_poly_nearhorizon}

In this subsection, our objective is to prove the general formula for the $\mu$-polynomial constraint $P_n(\mu) = 0$, specifically Eq. \eqref{equ_Pn_mu_linear_explicit}.

We begin by rewriting the polynomial constraint \eqref{equ_Pn_mu_linear_explicit} using Vieta’s formulas, replacing $E_n(\mu^m)$ with the corresponding ratios $\frac{v_{n,n-m}}{v_{n,n}}$. This leads to the equivalent identity:
\begin{equation}\label{equ_Pn_mu_linear_Vieta}
\frac{v_{n, n-1}}{v_{n, n}}-\binom{n+1}{3}\frac{v_{2,1}}{v_{2,2}}+\frac{40 \binom{n+2}{5}}{(n - 1)(n + 1)}\frac{v_{1,0}}{v_{1,1}}= 0,
\end{equation}
Our goal is to derive and verify this identity directly.

To proceed, we recast the master equation \eqref{equ_master_equation_General} into a more compact form:
\begin{equation}\label{equ_master_equation_concise}
A (r,\omega,\mu)\Psi (r)+B (r,\omega)\Psi'(r)+C (r)\Psi''(r)=0,
\end{equation}
where $A(r,\omega,\mu)$, $B(r,\omega)$ and $C(r)$ take the form 
\begin{equation}\label{equ_specific_form_ABC}
\begin{aligned}
&A (r,\omega,\mu)=-(V (r) + \frac{\mu}{r^{2}} + \frac{i d \omega}{r g_{vr}(r)}), \\
&B (r,\omega)= -\frac{2 i \omega}{g_{vr}(r)} + \frac{d g_{vv}(r)}{r g_{vr}(r)^{2}} - \frac{g_{vv}(r) g_{vr}^{\prime}(r)}{g_{vr}(r)^{3}} + \frac{g_{vv}^{\prime}(r)}{g_{vr}(r)^{2}}, \\
&C (r)= \frac{g_{vv}(r)}{g_{vr}(r)^{2}}.\\
\end{aligned}
\end{equation}
These coefficients can be expanded in a Taylor series around the horizon (which we take to be at $r = r_h$ in this subsection to maintain generality) as follows:

\begin{equation}\label{equ_ABC_expansion}
\begin{aligned}
&A (r,\omega,\mu)=A_0 (\omega,\mu)+A_1 (\omega,\mu)(r-r_h)+A_2 (\omega,\mu)(r-r_h)^2+\ldots,\\
&B (r,\omega)=B_0 (\omega)+B_1 (\omega)(r-r_h)+B_2 (\omega)(r-r_h)^2+\ldots,\\
&C (r)=C_1 (r-r_h)+C_2 (r-r_h)^2+\ldots
\end{aligned}
\end{equation}
This allows us to express each element within the matrix $\mathcal{M}^{(n)}(\omega_{n},\mu)$ in a concise manner:

\begin{equation}\label{equ_M_ij_expression}
\mathcal{M}^{(n)}(\omega_{n},\mu)_{ij}=A_{i-j}(\omega_{n},\mu)+(j-1) B_{i-j+1}(\omega_{n})+(j-2)(j-1) C_{i-j+2},
\end{equation}
where $ A_{n} $, $ B_{n} $, and $ C_{n} $ are set to zero for $ n < 0 $. We can further isolate the dependence on $ \mu $ from $A_{i-j}(\omega_{n},\mu)$, leading to:

\begin{equation}\label{equ_A_dependent_mu}
A_{i-j}(\omega_{n},\mu)=\frac{(-1)^{i-j+1} (i-j+1)}{r_{h}^{i-j+2}}\mu+A_{i-j}(\omega_{n}, 0).
\end{equation}
Combining Eq. \eqref{equ_M_ij_expression} with Eq. \eqref{equ_A_dependent_mu}, the expression for $ \mathcal{M}^{(n)}(\omega_{n},\mu) $ can be explicitly written as:

\begin{equation}\label{equ_M_with_ABC}
\small
\scriptsize
\mathcal{M}^{(n)}(\omega_{n},\mu) \equiv \begin{pmatrix}
-\frac{\mu}{r_h^2}+A_0\left (\omega _n, 0\right) & B_0\left (\omega _n\right) & \cdots & 0 \\
\frac{2 \mu }{r_h^3}+A_1\left (\omega _n, 0\right) & -\frac{\mu}{r_h^2}+A_0\left (\omega _{n}, 0\right)+B_1\left (\omega _{n}\right) & \cdots & 0 \\
\vdots & \vdots & \ddots & \vdots \\
\left (-1\right)^{n}\frac{n \mu }{r_{h}^{n+1}}+A_{n-1}\left (\omega _{n}, 0\right) & \left (-1\right)^{n-1}\frac{(n-1) \mu }{r_{h}^{n}}+A_{n-2}\left (\omega _{n}, 0\right)+B_{n-1}\left (\omega _{n}\right) & \cdots & F(\omega_n,\mu)
\end{pmatrix},
\normalsize
\end{equation}
where $F(\omega_n,\mu)$ takes the form 

\begin{equation}
    -\frac{\mu }{r_h^2}+A_0\left (\omega _{n}, 0\right)+(n-1) B_1\left (\omega _{n}\right)+(n-2)(n-1) C_{n-2}.
\end{equation}
Also note that the $\mu$-terms only appear in the lower triangular part. 

Calculating the determinant of $\mathcal{M}^{(n)}(\omega_{n}, \mu)$ from Eq. \eqref{equ_M_with_ABC} yields a polynomial in $\mu$ whose highest degree is $n$. This is guaranteed by the fact that each diagonal entry on the right-hand side of Eq. \eqref{equ_M_with_ABC} contains a $\mu$-dependent term. By comparing this polynomial with the expression in Eq. \eqref{equ_mu_order_n_vieta}, we can extract the general formulas for $V_{n,n}$ and $V_{n,n-1}$ in Eq. \eqref{equ_mu_order_n_vieta} as follows:

\begin{equation}\label{equ_explicit_ann_annminus1}
\begin{aligned}
&V_{n, n}=(-1)^{n}r_{h}^{-2n},\\
&V_{n, n-1} = (-1)^{n+1} \Bigg ( \frac{2}{3} (n-2) (n-1) n r_{h}^{1-2n}C_{1} + \frac{1}{3} (n-2) (n-1) n r_{h}^{2-2n}C_{2} \\
& \qquad \qquad + n r_{h}^{2-2n}A_{0}(\omega_{n}, 0) + n (n-1) r_{h}^{1-2n}B_{0}(\omega_{n}) + \frac{1}{2}n (n-1) r_{h}^{2-2n}B_{1}(\omega_{n}) \Bigg),\\
\end{aligned}
\end{equation}
which relate to $v_{n,m}$ through $v_{n,m} = (-1)^{n - m} V_{n,m}$. The explicit expressions for the near-horizon expansion coefficients $A_{0}(\omega_{n}, 0)$, $B_{0}(\omega_{n})$, $B_{1}(\omega_{n})$, $C_{1}$, and $C_{2}$ can be readily derived from Eq. \eqref{equ_specific_form_ABC} as:

\begin{equation}\label{equ_ABC_explicit}
\hspace*{-1.4cm}
\begin{aligned}
&A_{0}(\omega_{n}, 0)= -\frac{d g_{vv_1} n}{2 g_{vr_0}^2 r_h} - V_0, \quad B_{0}(\omega_{n}) = (1-n)\frac{g_{vv_1}}{g_{vr_0}^2}, \\
&B_{1}(\omega_{n}) = \frac{d g_{vv_1}}{g_{vr_0}^2 r_h} + \frac{g_{vr_1} g_{vv_1} n}{g_{vr_0}^3} - \frac{3 g_{vr_1} g_{vv_1}}{g_{vr_0}^3} + \frac{2 g_{vv_2}}{g_{vr_0}^2}, \\
&C_{1} = \frac{g_{vv_1}}{g_{vr_0}^2}, \quad C_{2}= \frac{g_{vv_2}}{g_{vr_0}^2} - \frac{2 g_{vr_1} g_{vv_1}}{g_{vr_0}^3}.\\
\end{aligned}
\end{equation}
Substituting both Eq. \eqref{equ_explicit_ann_annminus1} and Eq. \eqref{equ_ABC_explicit} into the LHS of Eq. \eqref{equ_Pn_mu_linear_Vieta} results in the cancellation of all terms, thereby proving the validity of the identity \eqref{equ_Pn_mu_linear_Vieta} and confirming the original linear $\mu$-constraint \eqref{equ_Pn_mu_linear_explicit}.

This completes the proof of the linear $\mu$-constraint \eqref{equ_Pn_mu_linear_explicit}, achieved by exploiting the special algebraic structure of $\text{Det}(\mathcal{M}^{(n)}(\boldsymbol{\mu})) = 0$. This structure originates from the near-horizon expansion of the Klein-Gordon-type master equation \eqref{equ_master_equation_General} evaluated at pole-skipping points. It is natural to expect that the proof strategy for more general constraints $P_n(\mu^m)=0$ with $m > 1$ may follow similar lines, although the explicit expressions for $V_{n,\, n-m}$ in terms of $A_i(\omega_n,0)$, $B_i(\omega_{n})$, and $C_i$ are likely to be significantly more intricate.

\section{$\mu$-polynomial constraints in several examples}\label{sec_example_verify_polynomial_constraint}
To validate the universality of these $\mu$-polynomial constraints, we examine their validity across various holographic models whose master equations satisfy Assumptions \ref{ass_1}.

\subsection{Probe scalar field perturbations in four-dimensional Lifshitz black holes}

We begin by studying a probe scalar field of mass $m$ in a four-dimensional Lifshitz black hole background, as proposed in \cite{balasubramanian_2009_analytic_Lif_black_hole}. The background geometry is described by the metric \eqref{equ_hyperscaling_metric}, where the dynamical exponent is set to $\mathbf{z}=2$, and the hyperscaling violation exponent $\theta$ is turned off in Eq. \eqref{equ_hyper_violation_scaling}. Under this scaling, the components $g_{vv}(r)$ and $g_{vr}(r)$ take the form:

\begin{equation} \label{equ_Lif_black_hole} g_{vv}(r) = r^{2} - r^{4}, \quad g_{vr}(r) = r, 
\end{equation}
where we have set $r_{h} = 1$ by exploiting the scaling symmetry in Eq. \eqref{equ_hyper_violation_scaling_KGequation}. According to the holographic dictionary, the probe scalar field corresponds to a boundary operator of dimension $\Delta$, with its mass related by $m^2 = \Delta(\Delta - 4)$. The associated Klein-Gordon equation for the Fourier mode $\varphi$ of this scalar field is given by

\begin{equation} \label{equ_KGequation_Lif} 
\left( -\frac{k^2 + m^2 r^2 + 2 i \omega}{r^2} \right) \varphi(r) + \left( \frac{-3 + 5 r^2 - 2 i \omega}{r} \right) \varphi'(r) + \left(-1 + r^2 \right) \varphi''(r) = 0.
\end{equation}
The corresponding pole-skipping points can be derived via near-horizon analysis, yielding

\begin{equation} \label{equ_PS_points_Lif} \omega_{n} = -i n, \quad \mu_{n, q} = 2n(\Delta + 2q - 3) - (\Delta + 2q - 4)(\Delta + 2q - 2), \end{equation}
where we have used $T = \frac{1}{2\pi}$ and $m^2 = \Delta(\Delta - 4)$, with $q$ ranging from $1$ to $n$. The pole-skipping locations in \eqref{equ_PS_points_Lif} clearly satisfy Assumptions \ref{ass_1}. Therefore, we expect the pole-skipping points \eqref{equ_PS_points_Lif} to satisfy the $\mu$-polynomial constraints $P_{n}(\mu) = 0$ \eqref{equ_Pn_mu_linear_explicit}, $P_{n}(\mu^2) = 0$ \eqref{equ_Pn_mu_2_explicit}, $P_{n}(\mu^{3})=0$ \eqref{equ_Pn_mu_3_explicit} and $P_{n}(\mu^{4})=0$ \eqref{equ_Pn_mu_4_explicit}, all of which can indeed be confirmed with ease. This example indicates that the validity of the $\mu$-polynomial constraints is unaffected by the asymptotic boundary behavior of the background metric, and thus extends beyond asymptotically AdS spacetimes, as we previously claimed.

\subsection{$U(1)$ gauge field perturbations in $d+2$-dimensional AdS black holes}

We then consider the $U(1)$ gauge field $A_{\mu}$ in a $d+2$-dimensional asymptotic AdS spacetime, which is described by the metric in \eqref{equ_general_metric_d}, with the components

\begin{equation}\label{equ_gvv_gvr_AdS_d}
g_{vv} = r^{2}-\left( \frac{1}{r} \right)^{d-1}, \quad g_{vr} = 1, 
\end{equation}
where we have set $r_h = 1$. According to the holographic dictionary, this bulk $U(1)$ gauge field $A_{\mu}$ is dual to the boundary conserved $U(1)$ charge current operator $J^{\mu}$. 

The corresponding bulk Maxwell equation is given by $\nabla^{\mu} F_{\mu\nu}=0$, where $F_{\mu\nu} = \nabla_\mu A_\nu - \nabla_\nu A_\mu$. Now, considering gauge field perturbations $\delta A_{\mu}$, we align the momentum direction along the $x$-axis, so that the perturbation fields $\delta A_{\mu}$ can be divided into transverse and longitudinal sectors, depending on whether they are perpendicular or parallel to the momentum direction. Here, we focus on the longitudinal perturbations, which are governed by a single Klein-Gordon equation:

\begin{equation}\label{equ_longitudinal_Maxwell_d}
\small
\begin{aligned}
&\Bigg(
\frac{(d-1)^2}{r^{d+1}}+
\frac{ 4 d r (r - i \omega)-4\mu}{r^2} -
\frac{(1 + d) \left((-3 + d) \omega^2-(1 + d)\mu \right)}{\mu - r^{1 + d} (\mu - \omega^2)}-\frac{3 (1 + d)^2 \mu \omega^2}{\left( \mu + r^{1 + d} (\omega^2-\mu) \right)^2}\Bigg)\Phi(r)\\
&+4\Big((d+2) r-2 i \omega-r^{-d}\Big)\Phi'(r)+4r \big(r - r^{-d}\big)\Phi''(r)=0,\\
\end{aligned}
\end{equation}
where $\Phi=\omega\delta A_{x}+k\delta A_{v}$. The structure of the pole-skipping points for this equation was analyzed in \cite{blake_2020_lower_infinite_pole_skipping_1}, and it satisfies Assumption \ref{ass_1}. The explicit expressions for $E_{n}(\mu)$ and $E_{n}(\mu^2)$ with $n < 5$ are given by:
\begin{equation}\label{equ_PS_data_Maxwell}
\begin{aligned}
&\mkern10mu E_{1}(\mu)=\frac{1}{2} (d-2) (d+1),\quad E_{2}(\mu)=-2 (d+1),\quad E_{3}(\mu)=-\frac{1}{2} (d+1) (5 d+6),\\
&E_{4}(\mu)=-4 (d+1) (2 d+1),\quad E_{2}(\mu^{2})=(d-2) (d-1) (d+1)^2,\\
&E_{3}(\mu^{2})=-\frac{1}{4} (d+1)^2 \left (9 d^2-64 d+36\right),\mkern5mu  E_{4}(\mu^{2})=12 (d+1)^2 \left (d^2+4 d-2\right),\\
\end{aligned}
\end{equation}
which constitute the previously derived polynomial constraints: $P_3(\mu)= 0$ \eqref{equ_linear_mu_constrain_3}, $P_4(\mu) = 0$ \eqref{equ_P4_mu_linear}, and $P_4(\mu^2) = 0$ \eqref{equ_P4_mu_quadratic_invariant}. Substituting the data from \eqref{equ_PS_data_Maxwell} into these constraints will readily confirm their validity.

This example confirms that the $\mu$-polynomial constraints are valid, regardless of the dimension of the background spacetime, and also apply to gauge field perturbations.

\subsection{Metric perturbations in four-dimensional massive black holes}\label{subsubsec_4d_massive_gravity}

We conclude with the case of pure gravitational perturbations in four-dimensional static maximally symmetric black holes, as studied in \cite{grozdanov_2023_Pole_skipping_hyperbolic_sphereical_flat}. The relevant background metric is given by
\begin{equation}\label{equ_maximally_symmetric_4d}
ds^2=-g_{vv}(r) dv^2 + 2g_{vr}(r) dvdr+r^{2}\gamma_{ab}dx^{a}dx^{b},
\end{equation}
where $\gamma_{ab}dx^{a}dx^{b}$ denotes a $2d$ maximally symmetric Riemannian space, which can be written in the general form 
\begin{equation}\label{equ_Riemannian_space}
\gamma_{ab}dx^{a}dx^{b}=\frac{ d \chi^2}{1-K \chi^2}+\chi^2 d \phi,
\end{equation}
and the $g_{vv}(r),\mkern6mu g_{vr}(r)$ components take the form 
\begin{equation}\label{equ_gvv_gvr_4d_gravity}
g_{vv}(r)=K-\frac{2M}{r}+\frac{r^{2}(2M-K r_{h})}{r_{h}^{3}}, \quad  g_{vr}(r)=1,
\end{equation}
with $M$ being a constant representing the mass of black holes. The parameter $K$ in Eqs. \eqref{equ_Riemannian_space} and \eqref{equ_gvv_gvr_4d_gravity} denotes the normalized sectional curvature and takes the values $K = 1$, $K = 0$, or $K = -1$, corresponding to spherical, planar, or hyperbolic horizon, respectively. Following the conventions of \cite{grozdanov_2023_Pole_skipping_hyperbolic_sphereical_flat}, we also define the normalised temperature as
\begin{equation}\label{equ_definition_norm_tem}
\tau=\frac{4M (3M-K r_{h})}{r_{h}^{2}}. 
\end{equation}
In the metric \eqref{equ_maximally_symmetric_4d}, different choices of $K$ and $\tau$ parameterize asymptotically Minkowski, de Sitter, or anti-de Sitter black holes with flat, spherical, or hyperbolic horizons in four-dimensional. For a detailed classification, we refer the reader to \cite{grozdanov_2023_Pole_skipping_hyperbolic_sphereical_flat}.

In this setting, all metric perturbations $\delta g_{\mu \nu}$ decompose into parity-odd ($-$) and parity-even ($+$) sectors, each governed by decoupled master equations $(\nabla^2 + V_\pm(r))\Phi_\pm(r) = 0$, where $\Phi_\pm(r)$ represent the master fields for odd ($-$) and even ($+$) sectors. We focus on the odd sector, where $V_-(r) =\frac{6M-\mu r}{r^3}$ and the associated master equation becomes

\begin{equation}\label{equ_4d_odd_master_equation}
\begin{aligned}
&\frac{ (6 M-\mu  r)}{r^3}\Phi_{-} (r)+\left (\frac{2 M}{r^2}+\frac{2r(2M-K r_h)}{r_h^3}-2 i \omega \right)\Phi_{-}'(r)\\
&+\left(K-\frac{2M}{r}+\frac{r^{2}(2M-K r_{h})}{r_{h}^{3}}\right)\Phi_{-}''(r)=0.\\
\end{aligned}
\end{equation}
Note that $V_-(r)$ being linear in $\mu$ ensures that the determinant equation \mbox{$\text{Det}(\mathcal{M}^{(n)}(\boldsymbol{\mu}))=0$}, derived from \eqref{equ_4d_odd_master_equation}, yields $n$ roots $\mu_{n,q}$ for any $n$, implying that Assumption \ref{ass_1} holds here.

As detailed in \cite{grozdanov_2023_Pole_skipping_hyperbolic_sphereical_flat}, the Darboux transformation formalism allows analytical determination of the ``algebraically special'' (AS) pole-skipping points for both odd and even sectors without implementing the near-horizon analysis. For the odd sector, these AS points are:\footnote{At $n=0$, there is another AS point $\mu=2K$.}
\begin{equation}\label{equ_AS_pole_skipping_odd}
\begin{aligned}
&n=1:\quad\mu=K+\sqrt{K^2+3\tau},\\
&n\geq2:\quad\mu=K\pm\sqrt{K^2+3n\tau}.\\
\end{aligned}
\end{equation}
For each $n > 2$, the remaining $n-2$ ``common'' pole-skipping points, shared by both sectors, can be determined by solving the corresponding determinant equation $\text{Det}(\mathcal{M}^{(n)}(\boldsymbol{\mu})) =0$. Alternatively, these points can also be identified by solving the set of $n-2$ $\mu$-polynomial constraints $P_n(\mu^m) = 0$.\footnote{Although, strictly speaking, these identities are derived under the planar case ($K=0$).}

To illustrate this explicitly, we set the AS pole-skipping points as $\mu_{n,1}$ and $\mu_{n,2}$ for any $n > 2$, then compare the values of the remaining $\mu_{n,q}$ obtained from solving $\text{Det}(\mathcal{M}^{(n)}(\boldsymbol{\mu})) =0$ and from exploiting $P_{n}(\mu^{m})=0$. 

At $n = 3$, combining the AS pole-skipping points from Eq. \eqref{equ_AS_pole_skipping_odd} with $P_{3}(\mu) = 0$ \eqref{equ_linear_mu_constrain_3} yields $\mu_{3,3} = K-5 \sqrt{K^2+3 \tau}$, which can be easily verified as consistent with the result obtained by solving $\text{Det}(\mathcal{M}^{(3)}(\boldsymbol{\mu}))= 0$.

For $n = 4$, substituting the solution for $\mu_{3,3}$ along with all AS pole-skipping points up to $n = 4$ into $P_{4}(\mu) = 0$ \eqref{equ_P4_mu_linear} and $P_{4}(\mu^2) = 0$ \eqref{equ_P4_mu_quadratic_invariant} allows us to solve for $\mu_{4,3}$ and $\mu_{4,4}$ (up to a permutation) as:

\begin{equation}\label{equ_4d_odd_mu43_mu44} \mu_{4,3} = -8 \sqrt{K^2+3 \tau }-\sqrt{9 K^2+12 \tau }+K, \quad \mu_{4,4} = -8 \sqrt{K^2+3 \tau }+\sqrt{9 K^2+12 \tau }+K. 
\end{equation}
Again, solution \eqref{equ_4d_odd_mu43_mu44} is consistent with those obtained by solving $\text{Det}(\mathcal{M}^{(4)}(\boldsymbol{\mu}))= 0$.

\begin{figure}[h!]
\centering
 \includegraphics[width=1\textwidth]{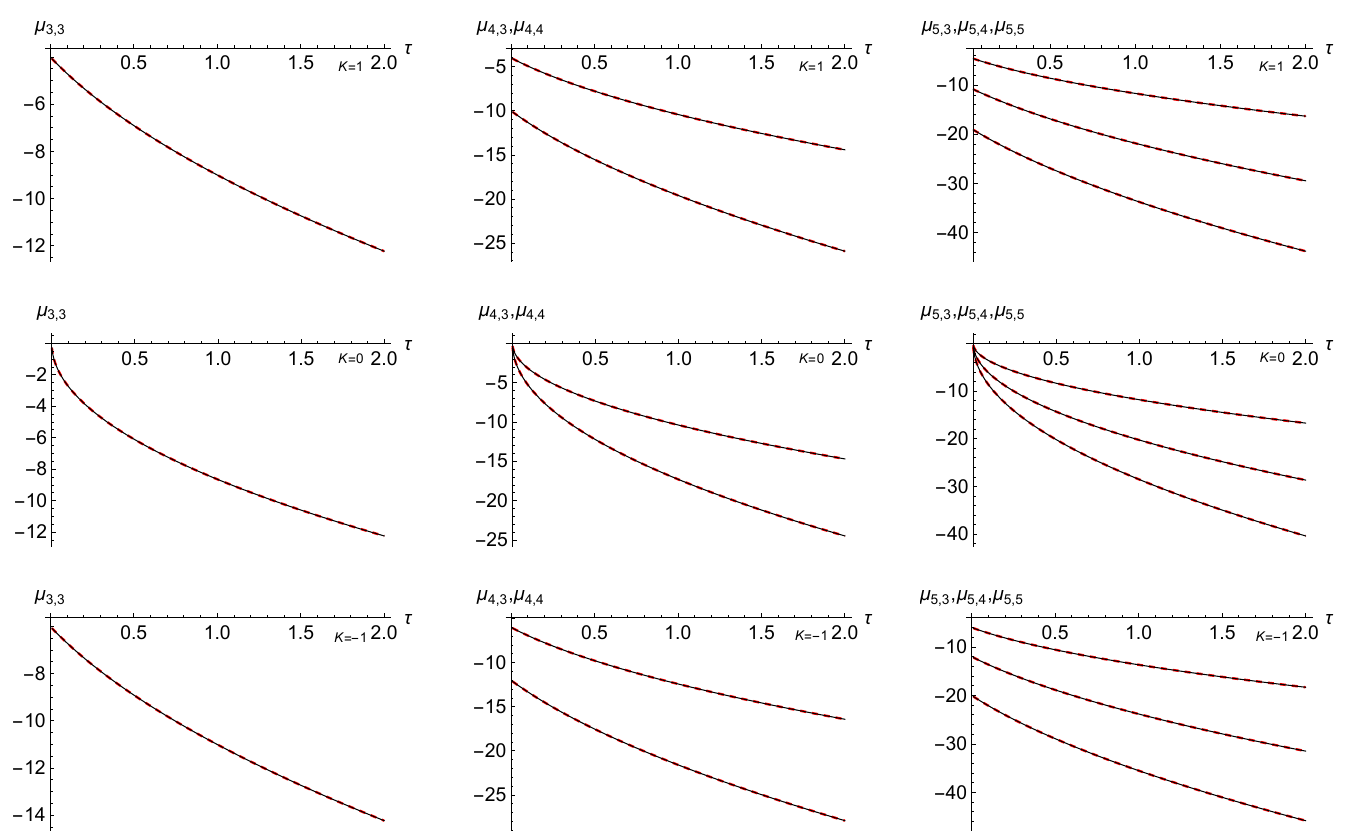}
\caption{\label{fig_poly_deter_4d_odd}For each row, from left to right, the figure compares the ``common'' pole-skipping points $\mu_{n,q}$ obtained by solving $\text{Det}(\mathcal{M}^{(n)}(\boldsymbol{\mu}))  = 0$ (depicted by the red dashed line) with those determined by incorporating $P_{n}(\mu^{m})= 0$ (depicted by the black solid line) for $n = 3, 4, 5$, spanning the full parameter space of $K$ and $\tau$. From top to bottom, the rows correspond to $K = 1$, $0$, and $-1$, respectively.}
\end{figure}

The same procedure applies for larger $n$: Given the AS pole-skipping points in Eq. \eqref{equ_AS_pole_skipping_odd}, the remaining $n-2$ pole-skipping points can be determined either by solving $\text{Det}(\mathcal{M}^{(n)}(\boldsymbol{\mu}))  = 0$ or by using all polynomial constraints $P_{n}(\mu^{n-2}) = 0,\mkern6mu \ldots,\mkern6mu P_{n}(\mu) = 0$. The consistency between these two methods is explicitly illustrated in Figure \ref{fig_poly_deter_4d_odd}.

As shown in Figure \ref{fig_poly_deter_4d_odd}, the $\mu$-polynomial constraints hold even in spherical and hyperbolic black holes (at least in four dimensions), and indeed hold despite the differences in boundary asymptotic behavior. Furthermore, we can conclude that in four-dimensional maximally symmetric black holes, the AS points derived from the Darboux transformation formalism, combined with the implementation of the $\mu$-polynomial constraints, are sufficient to determine \textbf{all} the pole-skipping points of both sectors. 

In this section, we have verified the validity of the $\mu$-polynomial constraints across a broad range of holographic scenarios, encompassing different perturbation modes and master equations. However, we did not examine perturbations in the scalar sector associated with the boundary energy density Green's functions, which feature quantum chaos-related pole-skipping points at $(\omega_{c}, \mu_{c})$. This omission stems from our observation that Assumption \ref{ass_1} is always violated in these cases. Nonetheless, there are indications that these chaotic pole-skipping points can be incorporated into a `modified' form of the $\mu$-polynomial constraints, which we plan to explore in future work. 


\section{Discussion}\label{sec_discussion}

In this paper, we develop a novel boundary-to-bulk map that enables an analytic reformulation of the near-horizon expansion coefficients of the metric components: $g_{vv_{n}}$ and $g_{vr_{n-1}}$ in terms of boundary pole-skipping data, the frequency $\omega_{1}$ and elementary symmetric polynomials $E_{n}(\mu^{m})$, by solving a system of linear equations. This mapping further allows all geometric quantities dependent on the metric, including the near-horizon expansion of the vacuum Einstein equations, to be reinterpreted directly in terms of the same pole-skipping data.

By computing $g_{vv_n}$ and $g_{vr_{n-1}}$ to sufficiently high order, one can approximate the metric functions $g_{vv}(r)$ and $g_{vr}(r)$ with arbitrary accuracy within the convergence radius, determined by the nearest singularities in the complexified $r$-plane. If this convergence domain extends to the boundary, the reconstructed profiles can be used to solve the Klein-Gordon equation \eqref{equ_KGequation_General_d} and, following the standard holographic procedure, compute the retarded Green's function $\mathcal{G}^{\mathcal{O}}_{R}(\omega,k)$ (typically via numerical methods). Therefore, bypassing the intermediate step of reconstructing $g_{vv_{n}}$ and $g_{vr_{n-1}}$, the entire procedure effectively reconstructs $\mathcal{G}^{\mathcal{O}}_{R}(\omega,k)$ directly from the infinite set of discrete pole-skipping points $(\omega_n, \mu_{n,i})$, consistent with the findings of \cite{grozdanov_2023_Pole_skipping_reconstruct_spectrum}, though achieved via a different approach.


An interesting question is how the boundary-to-bulk map is affected when the bulk geometry is modified. As discussed in Section \ref{subsec_TTbar}, a particularly instructive setting is the $T\bar{T}$ deformation of the boundary CFT \cite{cavaglia_2016_TbarT-deformation_original_2a, smirnov_2017_TbarT-Deformation_original_1}. Holographically, this deformation corresponds to introducing a finite radial cutoff in the bulk spacetime \cite{mcgough_2018_TbarT-deformation_Holographya}, effectively removing the UV (near-boundary) region. As illustrated in Fig. \ref{fig_TTbar_GR}, this truncation of the near-boundary region causes the poles and zeros of the Green’s function to blur together, making higher-order pole-skipping points increasingly difficult to distinguish in practice.\footnote{We anticipate similar blurring effects to occur in other scenarios where the bulk geometry is modified.} This effect becomes especially apparent in the zoomed-in view of Fig. \ref{fig_TTbar_GR_enlarge}, where the pole-skipping locations are only discernible under very high resolution.


A key by-product of our reconstruction method is the set of $\mu$-polynomial constraints, namely $P_{n}(\mu^{n-2})=0$, $P_{n}(\mu^{n-3})=0$, $\ldots$, $P_{n}(\mu^{2})=0$ \eqref{equ_Pn_mu_2_explicit} and $P_{n}(\mu)=0$ \eqref{equ_Pn_mu_linear_explicit}, for any $n>2$, where $n$ denotes the order of the near-horizon expansion. These constraints indicate that among the $n$ pole-skipping points $(\omega_n, \mu_{n,q})$, only two are independent; the remaining $n-2$ points are fixed (up to a permutation) by these polynomial constraints. This redundancy guarantees that the number of independent $\mu_{n,q}$ matches the number of near-horizon expansion coefficients $g_{vr_{n-1}}$ and $g_{vv_n}$ for each $n > 2$. 

The existence of these $\mu$-polynomial constraints demonstrates that the bulk metric is redundantly encoded in the boundary pole-skipping points: for any $n > 2$, the near-horizon coefficients $g_{vv_{n}}$ and $g_{vr_{n-1}}$ can be reconstructed from any two out of the $n$ pole-skipping points, with the remaining $n - 2$ automatically constrained by these polynomial constraints. As illustrated in Section \ref{subsubsec_4d_massive_gravity}, in four-dimensional static black holes, the two independent pole-skipping points at each order $n$ naturally coincide with the algebraically special points, which can be determined analytically by Darboux transformation formalism \cite{grozdanov_2023_Pole_skipping_hyperbolic_sphereical_flat}. This analytical input then enables the complete set of pole-skipping points in both sectors to be derived by applying these $\mu$-polynomial constraints.

These polynomial constraints hold in a broad and general context. In this paper, we derive them and verify their validity for any master equation of the form $(\nabla^2 + V(r))\Phi(r) = 0$, where the pole-skipping structure satisfies Assumption \ref{ass_1} and $\nabla$ is defined on static, planar-symmetric black holes in arbitrary spacetime dimensions. We expect that these constraints remain applicable even in more general geometries, such as non-maximally symmetric spacetimes, as long as Assumption \ref{ass_1} continues to be satisfied.

We conclude with a discussion of several questions and directions for future research.

1. \textbf{Immediate generalizations} \newline
In this paper, our general arguments regarding bulk reconstruction and $\mu$-polynomial constraints are based on planar symmetric black holes and bosonic fields. It would be interesting to explore the applicability of our findings in broader contexts, including non-maximally symmetric black holes (such as rotating black holes), black holes with spherical or hyperbolic horizons, and black hole backgrounds coupled with fermionic fields.

2. \textbf{$\mu$-polynomial constraints under covariant expansion formalism} \newline 
The $\mu$-polynomial constraints identified in our paper pertain only to lower half-plane pole-skipping points at frequencies $\omega_{n} = -i n 2\pi T$, where $n$ is a positive integer. In theories with fields of spin $l$, the pole-skipping points occur at frequencies $\omega_{n} = i(l-n)2\pi T$, where $l$ can be either an integer (for bosons) or a half-integer (for fermions) \cite{wang_2022_Pole-skipping_gauge_bosonicfields, ning_2023_pole-skipping_gauge_fermionicfields}. 


As proposed in \cite{wang_2022_Pole-skipping_gauge_bosonicfields, ning_2023_pole-skipping_gauge_fermionicfields}, all pole-skipping points, including the chaos-related ones, can be  systematically derived using the covariant expansion formalism. This involves expanding fluctuations and equations of motion near the horizon using the covariant derivative $\nabla_{r}$, rather than partial derivatives $\partial_{r}$ as we did in our near-horizon expansion formalism. Under our current framework, the chaos-related pole-skipping points fall beyond the scope of the present $\mu$-polynomial constraints. It would be noteworthy to derive $\mu$-polynomial constraints using the covariant expansion formalism and explore the possibility of formulating more comprehensive $\mu$-polynomial constraints that encompass these chaos-related points.

3. \textbf{QFT foundations of the $\mu$-polynomial constraints} \newline
In Section \ref{sec_mu_polynomial_constraint}, we attribute the existence of these $\mu$-polynomial constraints to the unique algebraic structure underlying the near-horizon behavior of the Klein-Gordon type master equation \eqref{equ_master_equation_General} at pole-skipping points. While the emergence of these constraints does not depend on the existence of a boundary QFT, it would be highly intriguing to explore whether, and how, such constraints can be understood or derived from the perspective of a dual boundary QFT in cases where the bulk black hole admits a holographic interpretation.

4. \textbf{Prospects for observing Pole-Skipping Points in Simulation and Experiment}

Due to its analyticity and simplicity, our reconstruction method offers a promising framework that could inspire future experimental efforts to probe the emergence of spacetime, such as the search for spacetime-emergent materials \cite{koji_2022_space_emergent_material, koji_2025_ML_Green_function_reconstruct}. However, directly probing the complex-valued pole-skipping points remains an experimental challenge.

We outline several possible pathways to address this issue: 

\textbf{I.} Real-valued pole-skipping points have been identified in AdS solitons \cite{Natsuume_2023_pole-skipping_non-black_hole} and in holographic periodic lattice backgrounds \cite{balm_2020_Holography_fermion_lattice_potential}. In condensed matter systems, similar zero–pole interference phenomena, used to characterize the metal–Mott insulator transition \cite{sakai_2009_Mott_Simulator_poles_meet_zeros_simulation, sakai_2010_Mott_Simulator_poles_meet_zeros_simulation_2} or to distinguish topological insulators \cite{Misawa_2022_PS_detect_topological_insulator}, also occur at real frequency and momentum. It is thus intriguing to generalize and apply our reconstruction method to these contexts.

\textbf{II.} We anticipate that our reconstruction framework can be tested using lattice gauge theory \cite{Rothe_1992_Lattice_QCD_introduction_1, Gattringer_2010_Lattice_QCD_introduction_2} or quantum Monte Carlo simulation \cite{Foulkes_2001_QMC_introduction, Gull_2011_QMC_introduction}. Although such simulations typically yield Euclidean Green's functions, our method can be directly applied in Euclidean signature by substituting $\omega_{1} \to i\omega_{1}$ in the analytic expressions for $g_{vv_{n}}$ and $g_{vr_{n-1}}$. 

\textbf{III.} Numerical analytic continuation techniques (see, e.g., \cite{Jarrell_1996_analytic_continuation_introduction, Shao_2022_analytic_continuation_introduction_2}) can also be employed to extract the corresponding real-frequency spectral functions from the Euclidean Green's functions. Particularly, Ref. \cite{Huang_2025_barycentric_analytic_continuation} introduced an analytic continuation method based on barycentric rational functions, capable of locating the complex zeros and poles of Euclidean Green's functions, and thus may provide a practical route to identifying pole-skipping points. Along the same line, it is reasonable to anticipate that complex-valued pole-skipping points can be extracted from observable real-time Green's functions via analytic continuation into the complex frequency domain. 

\textbf{IV.} Ultimately, we anticipate that pole-skipping data may be directly extracted from quantum processor simulations of many-body systems. For a recent example of a quantum processor–based simulation of correlation functions, see \cite{kokcu_2024_Quantum_simulation_bosonfermion_correlation}.

\acknowledgments
    We would like to thank Navid Abbasi, Xian-Hui Ge, Song He, Yan Liu and Zhuo-Yu Xian for their helpful discussions. A special thanks goes to Sašo Grozdanov and Mile Vrbica for their frequent and insightful discussions, as well as their valuable feedback on the draft of this paper. SFW is supported by NSFC grants No.12275166 and No.12311540141.

\begin{appendix}

\section{Derive $m-\Delta$ relation via Pole-skipping points}\label{App_a_equal_1}

In this appendix, we aim to recover the $m$–$\Delta$ relation directly from the boundary pole-skipping data. We focus on the setup described in Section \ref{subsec_massive_reconstruction}, where $g_{vv_{1}}$ and $g_{vr_{0}}$ take the form:

\begin{equation}\label{equ_sol_gvv1_gvr0_d_mass}
g_{vv_1} = \frac{2 d r_{h} \omega_1^2}{m^2 r_{h}^{2}+ \mu_{1,1}}, \quad g_{vr_0} = -\frac{i d r_{h} \omega_1}{m^2 r_{h}^{2}+ \mu_{1,1}}.
\end{equation}
Compared to Eq. \eqref{equ_sol_gvv1_gvr0_d}, we have recast the horizon location as $r = r_h$ and denote $E_{1}(\mu)$ by $\mu_{1,1}$ to facilitate the subsequent analysis. To express $r_h$ and $m$ in terms of boundary quantities, we examine the common denominator shared by $g_{vr_0}$ and $g_{vv_1}$ in Eq. \eqref{equ_sol_gvv1_gvr0_d_mass}:

\begin{equation}\label{equ_dominator}
m^{2}r_{h}^{2}+\mu_{1,1}.
\end{equation}
We assume that the boundary QFT is invariant under the Lifshitz scaling transformation:

\begin{equation}\label{equ_scaling_Lifshitz}
\omega\to\omega/\lambda,\quad k\to k/\lambda^{1/\mathbf{z}},
\end{equation}
where $\mathbf{z}$ is the dynamical critical exponent. If $T_b$ is the only energy scale in the boundary theory, then dimensional analysis under the scaling \eqref{equ_scaling_Lifshitz} implies that $\mu_{1,1}$ must scale as $\mu_{1,1} = \hat{\mu}_{1,1} T_b^{2/\mathbf{z}}$, where $\hat{\mu}_{1,1}$ is dimensionless and independent of $T_b$. On the other hand, the mass term $m^2 r_h^2$ can be recast as $(\Delta^2 + b\Delta) r_h^2$ by examining the asymptotic boundary behavior of the Klein–Gordon equation, with the coefficient $b$ determined by the specific details of the theory.

In the probe limit, the expression \eqref{equ_dominator} must be independent of the scaling dimension $\Delta$, which implies that $\hat{\mu}_{1,1}$ must also take the quadratic form $\hat{\mu}_{1,1} = f_1 \Delta^2 + f_2 \Delta + f_3$, where $f_1$, $f_2$, and $f_3$ are $\Delta$-independent coefficients determined by the boundary QFT. Consequently, Eq. \eqref{equ_dominator} transforms to:

\begin{equation}\label{equ_dominator_Delta}
(\Delta^{2}+b\Delta) r_{h}^{2}+(f_{1}\Delta^{2}+f_{2}\Delta+f_3) T_{b}^{2/\mathbf{z}}.
\end{equation}
By canceling the $\Delta^{2}$ term, we derive $r_{h}^{2}=-f_{1}T_{b}^{2/\mathbf{z}}$, leading to $r_{h}=(-f_{1})^{1/2}T_{b}^{1/\mathbf{z}}$.\footnote{Given that both $r_{h}$ and $T_{b}$ are positive, the equation $r_{h}^{2}=-f_{1}T_{b}^{2/\mathbf{z}}$ necessitates $f_{1}<0$} Furthermore, we can express $b$ in terms of $f_{1}$ and $f_{2}$ as $b=\frac{f_{2}}{f_{1}}$ by eliminating the $\Delta$ term. This enables us to establish relationships between the bulk quantities and the boundary quantities: 

\begin{equation}\label{equ_T_b_Delta/m}
    r_h = (-f_{1})^{1/2} T_b^{1/\mathbf{z}}, \quad m^2 = \Delta^2+\frac{f_{2}}{f_{1}}\Delta.
\end{equation}


Extending the above demonstration to boundary QFT with multiple energy scales is straightforward. Specifically, assuming a dynamical exponent $z=1$ and considering a boundary QFT with two energy scales, denoted as temperature $T_{b}$ and chemical potential $\mu_{c}$, we can derive the $r_{h}-(T_{b},\mu_{c})$ relation and the values of $b$ in the $m-\Delta$ relation as $r_{h}=T_{b} (-f_{1}(\frac{\mu_{c}}{T_{b}}))^{1/2}$ and $b=\frac{f_{2}(\frac{\mu_{c}}{T_{b}})}{f_{1}(\frac{\mu_{c}}{T_{b}})}$. Here, $f_{1}$ and $f_{2}$ are no longer constants but functions of the dimensionless ratio $\frac{\mu_{c}}{T_{b}}$.

\section{Proof of the linearity of $\text{Det}(\mathcal{M}^{(n)}_{n-1}(\textbf{g}))$ and $\text{Det}(\mathcal{M}^{(n)}_{n}(\textbf{g}))$ with respect to $g_{vv_{n}}$ and $g_{vr_{n-1}}$}\label{App_linearity_prove}

In this appendix, we prove the statement: the last two equations within \mbox{$\text{Det}(\vec{\mathcal{M}}^{(n)}(\textbf{g}))=0$} derived from Klein-Gordon equation \eqref{equ_KGequation_General_d} is linear with respect to $g_{vv_{n}}$ and $g_{vr_{n-1}}$ after substituting all the solutions of $\text{Det}(\vec{\mathcal{M}}^{(m<n)}(\textbf{g}))=0$, i.e., $g_{vr_{0}}, \ldots, g_{vr_{n-2}}$ and $g_{vv_{1}}, \ldots, g_{vv_{n-1}}$ appearing within $\text{Det}(\vec{\mathcal{M}}^{(n)}(\textbf{g}))=0$ are treated as constants. Equivalently, we show that the terms $\frac{v_{n,1}}{v_{n,n}}$ and $\frac{v_{n,0}}{v_{n,n}}$ in Eq. \eqref{equ_last_two_equations_vieta} are linear in both $g_{vv_n}$ and $g_{vr_{n-1}}$.

To prove this statement, we adopt the notations used in Section \ref{subsec_explain_poly_nearhorizon} to reorganize the equation of motion \eqref{equ_KGequation_General_d} as:

\begin{equation}\label{equ_EOM_concise}
A (r,\omega,\mu)\varphi (r)+B (r,\omega)\varphi' (r)+C (r)\varphi'' (r)=0,
\end{equation}

To facilitate our proof, we multiply the original equation of motion \eqref{equ_KGequation_General_d} by an overall factor $g_{vr}(r)^{3}$. This introduces a prefactor of ${\left(g_{vr_{0}}\right)}^{3n}$ in front of the original \mbox{$\text{Det}(\mathcal{M}^{(n)}(\boldsymbol{\mu}))$}, which clearly does not affect the locations of the pole-skipping points $(\omega_{n}, \mu_{n,q})$ or the validity of our reconstruction method. With this extra factor, the new functions $A(r,\omega,\mu)$, $B(r,\omega)$, and $C(r)$ are given by

\begin{equation}\label{equ_specific_form_ABC_1}
\begin{aligned}
&A (r,\omega,\mu) = -\left(m^2 g_{vr}(r)^{3} + \frac{\mu}{r^{2}}g_{vr}(r)^{3} + \frac{i d \omega}{r}g_{vr}(r)^{2}\right), \\
&B (r,\omega) = -2 i \omega g_{vr}(r)^{2} + \frac{d g_{vv}(r) g_{vr}(r)}{r} - g_{vv}(r) g_{vr}^{\prime}(r) + g_{vv}^{\prime}(r) g_{vr}(r), \\
&C (r) = g_{vv}(r) g_{vr}(r). \\
\end{aligned}
\end{equation}
These functions can be expanded around the horizon as follows:

\begin{equation}\label{equ_ABC_expansion_1}
\begin{aligned}
&A (r,\omega,\mu) = A_0 (\omega,\mu) + A_1 (\omega,\mu)(r-r_h) + A_2 (\omega,\mu)(r-r_h)^2 + \ldots, \\
&B (r,\omega) = B_0 (\omega) + B_1 (\omega)(r-r_h) + B_2 (\omega)(r-r_h)^2 + \ldots, \\
&C (r) = C_1 (r-r_h) + C_2 (r-r_h)^2 + \ldots
\end{aligned}
\end{equation}

Analogous to Section \ref{subsec_explain_poly_nearhorizon}, each element of $\mathcal{M}^{(n)}(\boldsymbol{\mu})$ (associated with Eq. \eqref{equ_EOM_concise} and Eq. \eqref{equ_specific_form_ABC_1}) can be expressed in terms of the expansion coefficients of $A (r,\omega,\mu)$, $B (r,\omega)$, and $C(r)$:  

\begin{equation}\label{equ_M_ij_expression_1}
\mathcal{M}^{(n)}_{ij} = A_{i-j}(\omega_{n},\mu) + (j-1) B_{i-j+1}(\omega_{n}) + (j-2)(j-1) C_{i-j+2},
\end{equation}
where $\mathcal{M}^{(n)}_{ij}$ represents an element of $\mathcal{M}^{(n)}(\boldsymbol{\mu})$ located at the $i^{th}$ row and $j^{th}$ column. Similiar to Eq. \eqref{equ_A_dependent_mu}, we can also isolate the dependence on $\mu$ from $A_{i-j}(\omega_{n},\mu)$, leading to:

\begin{equation}\label{equ_A_dependent_mu_aij}
A_{i-j}(\omega_{n},\mu)=a_{i-j}\mu+A_{i-j}(\omega_{n}, 0),
\end{equation}
where $a_{i-j}$ denotes a coefficient composed of $g_{vr_{m}}$ with $m \leq i-j$.

We then analyze which elements $\mathcal{M}^{(n)}_{ij}$ among $\mathcal{M}_{ij}$ contain $g_{vv_{n}}$ and $g_{vr_{n-1}}$. This involves examining the dependence of $A_{i-j}$, $B_{i-j+1}$, and $C_{i-j+2}$ on $g_{vv_{n}}$ and $g_{vr_{n-1}}$ according to Eq. \eqref{equ_M_ij_expression_1}.

We start by examining the dependence of $g_{vr_{n-1}}$ in $A_{n-1}$, which is the coefficient of $(r-r_{h})^{n-1}$ in the near-horizon expansion of $A(r,\omega,\mu)$. According to Eq. \eqref{equ_specific_form_ABC_1}, a term in $A(r,\omega,\mu)$ that contributes $g_{vr_{n-1}}$ to $A_{n-1}$ is $g_{vr}(r)^{3}$. The expansion coefficients of this term in front of $(r-r_{h})^{n-1}$ take the form $g_{vr_{l_{1}}}g_{vr_{l_{2}}}g_{vr_{l_{3}}}$ with $l_{1}+l_{2}+l_{3}=n-1$. Among all possible combinations of $l_1$, $l_{2}$, and $l_{3}$, $g_{vr_{n-1}}$ only appears in the term $g_{vr_{0}}^{2}g_{vr_{n-1}}$, making it linear in $g_{vr_{n-1}}$ since $g_{vr_{m<n-1}}$ are treated as constants. Terms with $g_{vr_{m}}$ where $m>n-1$ are impossible because $l_1$, $l_{2}$, and $l_{3}$ cannot be negative. Other possible terms in $A(r,\omega,\mu)$ that contribute $g_{vr_{n-1}}$ to $A_{n-1}$ are $\frac{g_{vr}(r)^{3}}{r^{2}}$ and $\frac{g_{vr}(r)^{2}}{r}$. Similar to our analysis for $g_{vr}(r)^{3}$, the only terms containing $g_{vr_{n-1}}$ in the $(n-1)^{th}$ expansion coefficients of $\frac{g_{vr}(r)^{3}}{r^{2}}$ and $\frac{g_{vr}(r)^{2}}{r}$ are $\frac{g_{vr_{0}}^{2}g_{vr_{n-1}}}{r_{h}^{2}}$ and $\frac{g_{vr_{0}}g_{vr_{n-1}}}{r_{h}}$ respectively (up to a constant factor). Furthermore, $g_{vr_{n-1}}$ is the highest possible near-horizon expansion coefficient of $g_{vr}(r)$ that may appear in the $(n-1)^{th}$ expansion coefficients of $\frac{g_{vr}(r)^{3}}{r^{2}}$ and $\frac{g_{vr}(r)^{2}}{r}$. Hence, $A_{n-1}$ is linear in $g_{vr_{n-1}}$ and contains no terms with $g_{vr_{m}}$ where $m>n-1$. 

Applying an analog derivation to $B_{n-1}$ and $C_{n}$, we can prove that both are linear in $g_{vv_{n}}$ and $g_{vr_{n-1}}$ and do not contain any terms involving $g_{vv_{m}}$ and $g_{vr_{m-1}}$ where $m>n$. 

From the above demonstration, we can analyze the dependence of $\mathcal{M}^{(n)}_{ij}$ on $g_{vv_{n}}$ and $g_{vr_{n-1}}$ according to Eq. \eqref{equ_M_ij_expression_1}. Specifically, for the first term $A_{i-j}(\omega_{n},\mu)$ to have linear $g_{vr_{n-1}}$ dependence, we require $i-j=n-1$, which is the maximum value for $i-j$ and can only be achieved by taking $i=n$ and $j=1$. Similiarly, the second term $(j-1) B_{i-j+1}(\omega_{n})$ contributes linear terms in $g_{vv_{n}}$ and $g_{vr_{n-1}}$ only when $i = n$ and $j = 2$, so that $i - j + 1 = n - 1$. Finally, the last term $(j-2)(j-1) C_{i-j+2}$ does not contribute any linear terms in $g_{vr_{n-1}}$ or $g_{vv_{n}}$ for any $i$ and $j$ because for $j\geq 3$ there is no solution for $i$ satisfying $i-j+2=n$. The combination of these analyses for terms $A_{i-j}$, $B_{i-j+1}$, and $C_{i-j+2}$ indicates that linear terms involving $g_{vr_{n-1}}$ or $g_{vv_{n}}$ only appear in the $n^{th}$ (last) row of $\mathcal{M}^{(n)}_{ij}$, while elements in the preceding rows only contain $g_{vr_{m-1}}$ or $g_{vv_{m}}$ with $m<n$, which are constants as per our assumption.

We then compute $\text{Det}(\mathcal{M}^{(n)}(\boldsymbol{\mu}))$ via cofactor expansion along the $n^{th}$ row, yielding

\begin{equation}\label{equ_cofactor_expansion}
    \text{Det}(\mathcal{M}^{(n)}(\boldsymbol{\mu}))=\sum^{n}\limits_{j=1}\mathcal{M}^{(n)}_{nj}M_{nj},
\end{equation}
where $M_{nj}$ are the corresponding cofactors. From earlier analysis, linear dependence on $g_{vr_{n-1}}$ and $g_{vv_n}$ appears only in $\mathcal{M}^{(n)}_{n1}$ and $\mathcal{M}^{(n)}_{n2}$, while each cofactor $M_{nj}$ is polynomial in $g_{vv_{m}}$ and $g_{vr_{m-1}}$ with $m<n$, i.e., constants. 

Noting that $\mathcal{M}^{(n)}_{ij} = 0$ for $i - j \leq -2$, we can explicitly write:
\begin{equation}\label{equ_Mn1_Mn2_expansion}
\mathcal{M}^{(n)}_{n1}M_{n1}=\mathcal{M}^{(n)}_{n1}\mathcal{M}^{(n)}_{12}\mathcal{M}^{(n)}_{23}\cdots\mathcal{M}^{(n)}_{n-1\,n}, \quad \mathcal{M}^{(n)}_{n2}M_{n2}=\mathcal{M}^{(n)}_{n2}\mathcal{M}^{(n)}_{11}\mathcal{M}^{(n)}_{23}\cdots\mathcal{M}^{(n)}_{n-1\,n}.
\end{equation}
According to Eq. \eqref{equ_M_ij_expression_1}, only elements $\mathcal{M}^{(n)}_{ij}$ with $i \geq j$ carry $\mu$-dependence; all other elements are independent of $\mu$. This allows us to extract \textbf{all} the $\mu$-dependence explicitly in the following form:
\begin{equation}\label{equ_Mn1_Mn2_expression_2}
\begin{aligned}
&\mathcal{M}^{(n)}_{n1}M_{n1}
=\bigl( a_{n-1}\mu + A_{n-1}(\omega_{n}, 0) \bigr)
\mathcal{M}^{(n)}_{12} \mathcal{M}^{(n)}_{23} \cdots \mathcal{M}^{(n)}_{n-1\,n}, \\
&\mathcal{M}^{(n)}_{n2}M_{n2}
=\bigl( a_{n-2}\mu + A_{n-2}(\omega_{n}, 0) + B_{n-1}(\omega_{n}) \bigr)
\bigl( a_{0}\mu + A_{0}(\omega_{n}, 0) \bigr)
\mathcal{M}^{(n)}_{23} \cdots \mathcal{M}^{(n)}_{n-1\,n}.
\end{aligned}
\end{equation}


\sloppy All remaining terms within cofactor expansion \eqref{equ_cofactor_expansion}, $\mathcal{M}^{(n)}_{nj} M_{nj}$ with $j > 2$, only involve $g_{vr_{m-1}}$ and $g_{vv_m}$ for $m < n$, and thus do not contribute any linear terms in $g_{vv_{n}}$ or $g_{vr_{n-1}}$. Combining this with Eq. \eqref{equ_Mn1_Mn2_expression_2}, we conclude that only $v_{n,0}$ and $v_{n,1}$, corresponding to the constant and linear-in-$\mu$ terms in Eq. \eqref{equ_mu_order_n_vieta} respectively, depend linearly on $g_{vv_{n}}$ and $g_{vr_{n-1}}$. Moreover, since $v_{n,n}$ depends solely on $r_h$ (see Eq. \eqref{equ_explicit_ann_annminus1}), we prove that both $E_n(\mu^{n-1}) - \frac{v_{n,1}}{v_{n,n}}$ and $E_n(\mu^n) - \frac{v_{n,0}}{v_{n,n}}$ are indeed linear in $g_{vv_n}$ and $g_{vr_{n-1}}$.

The same derivation applies directly to both the massless Klein-Gordon equations \eqref{equ_KGequation_General_d_mass_zero} and \eqref{equ_KGequation_hyperviolation_d}. It also extends to cases where the equation of motion takes the form of the master equation \eqref{equ_master_equation_General}. The only adjustment required is to replace the $m^2 g_{vr}(r)^3$ term in $A(r,\omega,\mu)$ with $V(r) g_{vr}(r)^3$, which does not affect the validity of the original derivation.

\section{Derivation of the scalar Green's function under $T\bar{T}$ deformation}\label{App_TTbar_GR}

In this appendix, we derive the scalar retarded Green's function $\mathcal{G}_{r_c}^{\mathcal{O}}$ at a finite cutoff $r=r_c$ using the extended holographic dictionary proposed in \cite{hartman_2019_TbarT-Deformation_correlation_function}. According to this dictionary, we have
\begin{align}\label{TTZ} \langle e^{\int J\mathcal{O}} \rangle_{\mathrm{EFT}} = \int_{r < r_c} D\phi \ e^{iS_{\mathrm{bulk}}[\phi]}.
\end{align}
In standard quantization, the boundary value of $\phi$ corresponds to a source $J$ for a scalar operator $\mathcal{O}$ of dimension $\Delta$ in the dual boundary EFT:

\begin{align} 
\phi(r_c, x) = r_c^{\Delta - d - 1} J. 
\end{align}
We consider a free, massive, and probe scalar field $\phi$ of mass $m$ in the bulk, which leads to the following action:

\begin{equation}\label{equ_action_bulk}
    \begin{aligned}
    &S_{\mathrm{bulk}}=S+S_{ct},\\   
    &S=-\frac{1}{2}\int d^{d+2}x\sqrt{-g}\left((\partial\phi)^2+m^2\phi^2\right),\\
    &S_{ct}=\frac{1}{2}\int_{r=r_c}d^{d+1}x\sqrt{-\gamma} \kappa \phi^2.\\
\end{aligned}
\end{equation}
where $\kappa = -\Delta + d + 1$ is the counterterm coefficient, and $\gamma$ denotes the determinant of induced metric on the boundary EFT, defined as $\gamma_{ij} = r_c^{-2} g_{ij}|_{r=r_c}$.

The on-shell action at any finite cutoff $r_c$ is obtained via the saddle-point approximation in the large-$N$ limit:

\begin{equation} S_c = \int_{r=r_c} \frac{d\omega d^{d}k}{(2\pi)^{d+1}} \left( -\frac{1}{2} \phi_c \Pi_c + \frac{1}{2} \sqrt{-\gamma} \kappa \phi_c^2 \right). \end{equation}
Here, $\phi_c$ is the solution to the Klein-Gordon equation with Dirichlet boundary conditions at $r=r_c$, and $\Pi_c$ is the conjugate momentum in the radial direction:

\begin{align} \Pi_c \equiv -\sqrt{-g} g^{rr} \partial_r \phi \bigg|_{\phi=\phi_c}. \end{align}
The Fourier-transformed retarded Green's function for the dual scalar operator $\mathcal{O}$ in the boundary EFT is then given by
\begin{align}\label{GTT}
    \mathcal{G}_{r_c}^{\mathcal{O}}=-\frac{\delta^2S_c}{\delta J\delta J}\bigg\vert_{J=0}=(r_c^{\Delta-d-1})^2\left(-\frac{\Pi_c}{\phi_c}+\sqrt{-\gamma}\kappa\right)\bigg\vert_{r=r_c}.
\end{align}

Substituting the BTZ black hole metric into the Klein-Gordon equation and imposing the Dirichlet boundary condition at $r=r_c$ enables us to solve for $\phi_c$ explicitly as

{\small
\begin{align}\label{phic}
\phi_c(k,\omega,z) \propto & \, \left(1 - z^2\right)^{-\frac{i \omega}{2}} 
\Bigg[ 
\frac{z^{\Delta_{-}} \Gamma(\Delta) \Gamma\left( \frac{\Delta_- + \Omega_-}{2} \right) 
\Gamma\left( \frac{\Delta_- - \Omega_+}{2} \right) 
\, _2F_1\left( \frac{\Delta_- - \Omega_+}{2}, \frac{\Delta_- + \Omega_-}{2}; \Delta_-; z^2 \right)}
{\Gamma(\Delta_-) \Gamma\left( \frac{\Delta + \Omega_-}{2} \right) \Gamma\left( \frac{\Delta - \Omega_+}{2} \right)} \nonumber \\
& - z^\Delta \, _2F_1\left( \frac{\Delta - \Omega_+}{2}, \frac{\Delta + \Omega_-}{2}; \Delta; z^2 \right) 
\Bigg],
\end{align}
}
where we have used the coordinate $z=1/r$ with $z_h=1$. $\Omega_{+/-}$ are defined as
\begin{equation}
    \Omega_{+}=i k+i \omega, \quad \Omega_{-}=i \omega-i k. 
\end{equation}

Finally, by substituting the solution in Eq. \eqref{phic} into Eq. \eqref{GTT}, we derive the explicit form of $\mathcal{G}_{z_c}^{\mathcal{O}}(k, \omega, z_c)$:

\begin{equation}\label{equ_TTbar_GR}
\small
\begin{aligned}
    & \mathcal{G}^{\mathcal{O}}_{z_c}(\omega,k,z_c)=z_c^{4-2\Delta} \Bigg\{ \left(z_c^2 - 1\right) \Bigg[ z_c^{2\Delta} 
    \Gamma\left( \frac{1}{2}\left(\Delta - \Omega_{-}\right) \right) 
    \Gamma\left( \frac{1}{2}\left(\Delta - \Omega_{+}\right) \right) \\
    & \bigg( z_c^2 \left(\Delta + \Omega_{-}\right)\left(\Delta + \Omega_{+}\right) 
    {}_2 \tilde{F}_1 \left( \frac{1}{2} \left( \Delta + \Omega_{-} + 2 \right), 
    \frac{1}{2} \left( \Delta + \Omega_{+} + 2 \right); \Delta + 1; z_c^2 \right) \\
    & +4(\Delta - 1) {}_2 \tilde{F}_1 \left( \frac{1}{2} \left( \Delta + \Omega_{-} \right), 
    \frac{1}{2} \left( \Delta + \Omega_{+} \right); \Delta; z_c^2 \right) \bigg) \\
    & - z_c^4 \left( \Delta - \Omega_{-} - 2 \right) \left( \Delta - \Omega_{+} - 2 \right)
    \Gamma \left( -\frac{\Delta}{2} - \frac{\Omega_{-}}{2} + 1 \right)
    \Gamma \left( -\frac{\Delta}{2} - \frac{\Omega_{+}}{2} + 1 \right) \\
    & \times {}_2 \tilde{F}_1 \left( \frac{1}{2} \left( -\Delta + \Omega_{-} + 4 \right), 
    \frac{1}{2} \left( -\Delta + \Omega_{+} + 4 \right); 3 - \Delta; z_c^2 \right) \Bigg] \\
    & \Bigg/ \Bigg[ 2 z_c^4 \Gamma\left( -\frac{\Delta}{2} - \frac{\Omega_{-}}{2} + 1 \right)
    \Gamma \left( -\frac{\Delta}{2} - \frac{\Omega_{+}}{2} + 1 \right) \\
    & \times {}_2 \tilde{F}_1 \left( \frac{1}{2} \left( -\Delta + \Omega_{-} + 2 \right), 
    \frac{1}{2} \left( -\Delta + \Omega_{+} + 2 \right); 2 - \Delta; z_c^2 \right) \\
    & - 2 z_c^{2\Delta + 2} \Gamma\left( \frac{1}{2} \left( \Delta - \Omega_{-} \right) \right)
    \Gamma \left( \frac{1}{2} \left( \Delta - \Omega_{+} \right) \right) 
    {}_2 \tilde{F}_1 \left( \frac{1}{2} \left( \Delta + \Omega_{-} \right), 
    \frac{1}{2} \left( \Delta + \Omega_{+} \right); \Delta; z_c^2 \right) \Bigg] \\
    & + \frac{(\Delta - 2)\left(\sqrt{1 - z_c^2} - 1\right)}{z_c^2} + \Delta 
    + \frac{1}{2}\left(-\Omega_{-} - \Omega_{+}\right) - 2 \Bigg\},
\end{aligned}
\end{equation}
where ${}_2 \tilde{F}_1$ refers to the regularized hypergeometric function and we have set $T_b = \frac{1}{2\pi}$ in Eq. \eqref{equ_TTbar_GR}. It is straightforward to verify that $\mathcal{G}^{\mathcal{O}}_{z_c}(\omega, k, 0)$ recovers the BTZ scalar Green's function $\mathcal{G}^{\mathcal{O}}_{R}(\omega, k)$ as given in Eq. \eqref{equ_BTZ_GR}.

\section{General formulas for $P_n(\mu^3)$ and $P_n(\mu^4)$}\label{App_Pnmu3_Pnmu4}

\begin{equation}\label{equ_Pn_mu_3_explicit}
\small
\begin{aligned}
& \frac{1}{189} (2 + n) (-876 - 652n + 175n^2 + 140n^3) \binom{n}{5} E_1(\mu)^3- \frac{5}{3} (-177 - 35n + 28n^2) \binom{n+2}{7} E_1(\mu)^2 E_2(\mu) \\
& + \frac{2}{3} (-228 - 91n + 35n^2) \binom{n+2}{7} E_1(\mu) E_2(\mu)^2 - \frac{1}{9} (93 + 112n + 35n^2) \binom{n+1}{7} E_2(\mu)^3 \\
& - 56 (3 + n) \binom{n+2}{7} E_1(\mu) E_2(\mu^2)+ 2 (3 + n) (3 + 2n) \binom{n+1}{6} E_2(\mu) E_2(\mu^2) \\
& + 2 (16 + 7n) \binom{n+2}{7} E_1(\mu) E_3(\mu^2)- (13 + 7n) \binom{n+2}{7} E_2(\mu) E_3(\mu^2)+ \frac{432 \binom{n+4}{9}}{(-3 + n)(3 + n)} E_3(\mu^3) \\
& - \binom{n+3}{7} E_4(\mu^3) + E_n(\mu^3) = 0.
\end{aligned}
\end{equation}

\begin{equation}\label{equ_Pn_mu_4_explicit}
\small
\begin{aligned}
& -\frac{1}{1512} (2 + n) (21924 + 24798 n - 2021 n^2 - 7247 n^3 - 280 n^4 + 560 n^5) \binom{n}{6} E_1(\mu)^4 \\
& + \frac{8}{27} (8676 + 4056 n - 1697 n^2 - 525 n^3 + 140 n^4) \binom{n+2}{8} E_1(\mu)^3 E_2(\mu) \\
& - \frac{2}{9} (12198 + 7471 n - 1382 n^2 - 819 n^3 + 140 n^4) \binom{n+2}{8} E_1(\mu)^2 E_2(\mu)^2 \\
& + \frac{8}{3} (-657 - 563 n - 63 n^2 + 35 n^3) \binom{n+2}{9} E_1(\mu) E_2(\mu)^3- \frac{1}{15} (9 + 5 n) (127 + 126 n + 35 n^2) \binom{n+1}{9} E_2(\mu)^4 \\
& + 56 (-83 - 15 n + 12 n^2) \binom{n+3}{9} E_1(\mu)^2 E_2(\mu^2)- 112 (-46 - 15 n + 6 n^2) \binom{n+3}{9} E_1(\mu) E_2(\mu) E_2(\mu^2) \\
& + \frac{1}{15} (3 + n) (-573 - 516 n - 56 n^2 + 35 n^3) \binom{n+1}{7} E_2(\mu)^2 E_2(\mu^2) \\
& - \frac{1}{45} (3 + n) (673 + 611 n + 126 n^2) \binom{n+1}{7} E_2(\mu^2)^2 + \frac{1}{3} (1332 + 819 n - 18 n^2 - 56 n^3) \binom{n+2}{8} E_1(\mu)^2 E_3(\mu^2) \\
& + \frac{8}{3} (-180 - 126 n - 6 n^2 + 7 n^3) \binom{n+2}{8} E_1(\mu) E_2(\mu) E_3(\mu^2) - \frac{6}{5} (157 + 144 n + 35 n^2) \binom{n+2}{9} E_2(\mu)^2 E_3(\mu^2) \\
& + \frac{168}{5} (7 + 3 n) \binom{n+3}{9} E_2(\mu^2) E_3(\mu^2) - \frac{1}{5} (44 + 34 n + 7 n^2) \binom{n+2}{8} E_3(\mu^2)^2 \\
& - 2 (4 + n) (19 + 8 n) \binom{n+2}{8} E_1(\mu) E_3(\mu^3) + 8 (2 + n) (4 + n) \binom{n+2}{8} E_2(\mu) E_3(\mu^3) \\
& + 8 (8 + 3 n) \binom{n+3}{9} E_1(\mu) E_4(\mu^3) - 4 (7 + 3 n) \binom{n+3}{9} E_2(\mu) E_4(\mu^3) + \frac{880 \binom{5+n}{11}}{(-4 + n) (4 + n)} E_4(\mu^4) \\
& - \binom{4+n}{9} E_5(\mu^4) + E_n(\mu^4) = 0 \\
\end{aligned}
\end{equation}

\end{appendix}

\bibliographystyle{JHEP}
\bibliography{PS.bib}

\end{document}